\documentclass[final,5p,times,twocolumn]{elsarticle}

\usepackage{lineno}
\usepackage{amsmath}
\usepackage{float}
\usepackage{comment}
\usepackage{color}
\usepackage{physics}
\usepackage{graphicx}
\usepackage{dcolumn}
\usepackage{bm}
\usepackage{todonotes}
\usepackage[colorlinks=true,bookmarks=false,citecolor=blue,linkcolor=blue,hyperfootnotes=true,urlcolor=blue]{hyperref}
\usepackage[acronym]{glossaries}

\def\seven{Sr$_3$Ir$_2$O$_7$}
\def\four{Sr$_2$IrO$_4$}
\def\LSMO{ La$_{2/3}$Sr$_{1/3}$MnO$_3$}
\def\jeff{$J_{\text{eff}}=1/2$}

\newacronym{soc}{SOC}{spin-orbit coupling}
\newacronym{pld}{PLD}{pulsed laser deposition}
\newacronym{mbe}{MBE}{molecular beam epitaxy}
\newacronym{htsc}{HTSC}{high temperature superconductivity}
\newacronym{fm}{FM}{ferromagnetic}
\newacronym{afm}{AFM}{antiferromagnetic}
\newacronym{dm}{DM}{Dzyaloshinskii-Moriya}
\newacronym{rixs}{RIXS}{resonant inelastic x-ray scattering}
\newacronym{arpes}{ARPES}{angle-resolved photoemission spectroscopy}
\newacronym{rp}{RP}{Ruddlesden-Popper}
\newacronym{ma}{MA}{magnetic anisotropy}
\newacronym{pma}{PMA}{perpendicular magnetic anisotropy}
\newacronym{xmcd}{XMCD}{X-ray magnetic circular dichroism}
\newacronym{xld}{XLD}{X-ray linear dichroism}
\newacronym{xas}{XAS}{X-ray absorption spectroscopy}
\newacronym{amr}{AMR}{anisotropic magnetoresistance}

\newacronym{ood}{OOD}{oxygen octahedral deformation}
\newacronym{oor}{OOR}{oxygen octahedral rotation}

\modulolinenumbers[5]

\journal{Journal of Physics and Chemistry of Solids}

\begin{document}

\begin{frontmatter}

\title{Novel spin-orbit coupling driven emergent states in iridate-based heterostructures}

\author[add1]{Lin Hao}
\ead{lhao3@utk.edu}
\author[add2]{D. Meyers}
\ead{dmeyers@bnl.gov}
\author[add2]{M. P. M. Dean}
\ead{mdean@bnl.gov}
\author[add1]{Jian Liu}
\ead{jianliu@utk.edu}

\address[add1]{Department of Physics and Astronomy, University of Tennessee, Knoxville, Tennessee 37996, USA}
\address[add2]{Condensed Matter Physics and Materials Science Department, Brookhaven National Laboratory, Upton, New York 11973, USA}

\begin{abstract}
Recent years have seen many examples of how the strong \gls*{soc} present in iridates can stabilize new emergent states that are difficult or impossible to realize in more conventional materials. In this review we outline a representative set of studies detailing how heterostructures based on \gls*{rp} and perovskite iridates can be used to access yet more novel physics. Beginning with a short synopsis of iridate thin film growth, the effects of the heterostructure morphology on the \gls*{rp} iridates including \four{} and  SrIrO$_3$ are discussed. Example studies explore the effects of epitaxial strain, laser-excitation to access transient states, topological semimetallicity in SrIrO$_3$,  2D magnetism in artificial \gls*{rp} iridates, and interfacial magnetic coupling between iridate and neighboring layers. Taken together, these works show the fantastic potential for controlled engineering of novel quantum phenomena in iridate heterostructures.

\end{abstract}

\begin{keyword}
Iridates\sep Heterostructures\sep Spin-Orbit Coupling
\end{keyword}

\end{frontmatter}

\section{Introduction} \label{intro}

Within the last several decades, transition metal perovskites have dominated a significant portion of condensed matter research efforts due to their myriad of fascinating properties derived from their strong electron-electron correlations, with \gls*{htsc} leading the charge \cite{lee2006doping, Scalapino2012, keimer2015quantum, Armitage_EdopedCuprateReview2010,Tsuei_PairCupratesReview2000,anderson_1997cupratetheory}. Traditionally, perovskites hosting $3d$ orbitals, where the electron-electron repulsion, crystal field splitting, and bandwidth are the dominant energy scales, have garnered the most attention due to their abundance and relative ease of synthesis \cite{Cao2017,galasso2013structure}. In contrast, perovskites with $4d$, $5d$ transition metal ions have only recently, within the last two decades, started to gain traction within the field \cite{Crawford1994_214mag,Cao2002_327structure,maeno1994superconductivity,Witczak_correlatedSOC2014}.  Particularly, parallels relating these materials to the $3d$ \gls*{htsc} materials and their importance for the field of topologically protected phases stoked interest in these materials \cite{Cao2017,Witczak_correlatedSOC2014,jackeli2009mott,kim2009phase,kim2008novel,Kim2012_327RIXS,gao_214SC_2015,yan2015electron,kim2016observation,Gretarsson2016_dop214RIXS,Mitchell2015_214, wang2011twisted, meng2014odd}.

Part of the justification for the lag of these materials relative to their $3d$ counterparts can be attributed to the expected trends as one moves down the periodic table. Electron-electron correlation, which is chiefly responsible for the myriad of exciting states found in $3d$ materials, is strongly reduced as one goes from $3d$ to $4d$ to $5d$ orbitals, owing to the larger spatial extension which effectively reduces how strongly two electrons interact within an orbital, while, concurrently, the bandwidth increases through stronger covalency with surrounding ions. Taken together, these trends were expected to push these materials towards weakly correlated metallic phases, with increased \gls*{soc} having only minor effects \cite{kim2008novel,Ryden_IrO2RuO21970, Meyers_competition2014}.

\begin{figure}[t]
\includegraphics[width=1.0\columnwidth]{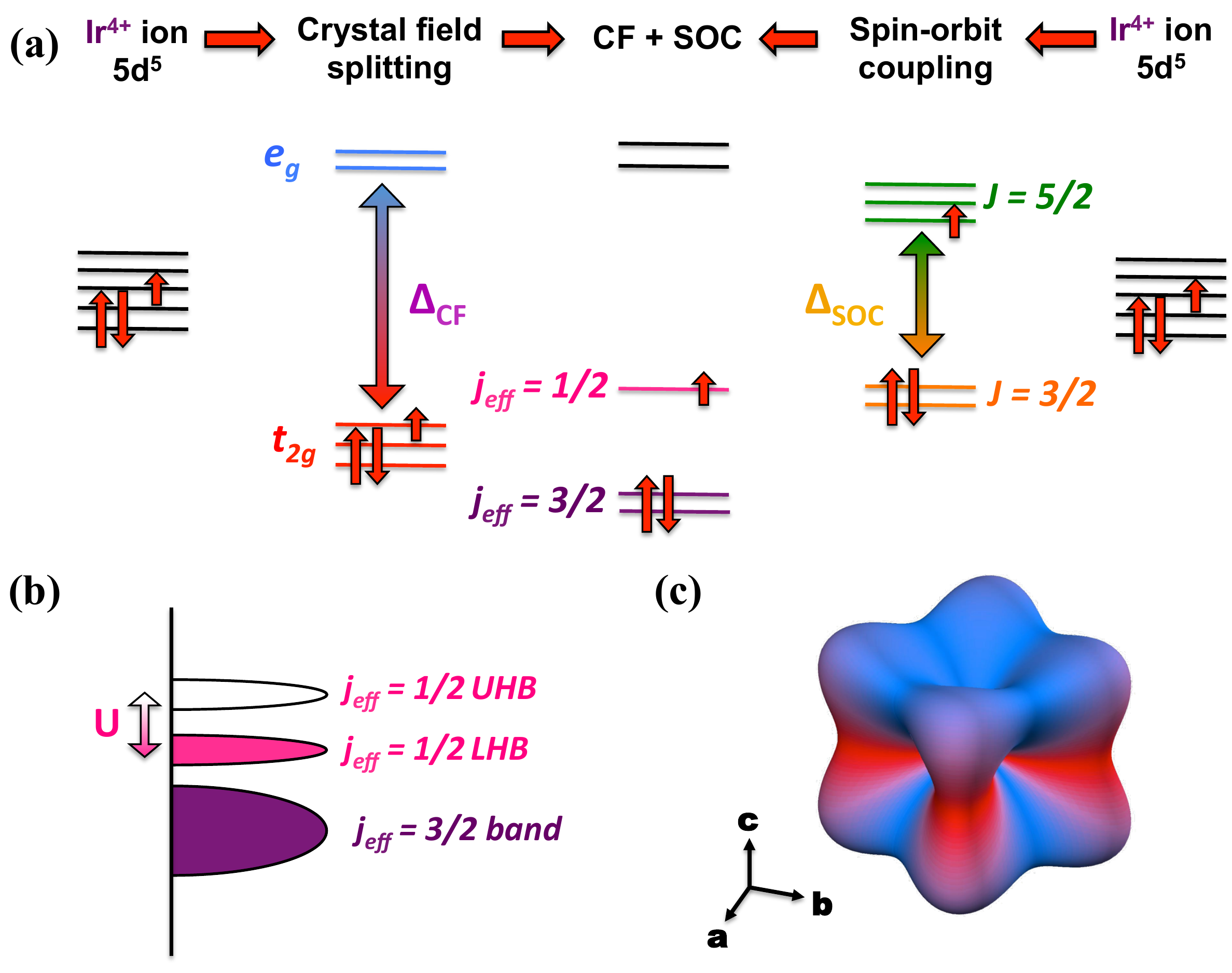}
\caption{\label{Jeff} (a) Cooperative effects of both crystal field splitting and \gls*{soc} on Ir$^{4+}$ (5$d^5$) ion in $O_h$ symmetry oxygen octahedra. (b) The final, singly occupied \jeff{} state is then split by electron-electron interactions to form an insulating state \cite{kim2008novel}. (c) \jeff{} orbital density profile for isospin up state \cite{jackeli2009mott}.
}
\end{figure}

Despite this, pioneering work exposed the error of these assumptions, showing, for instance, antiferromagnetic insulating states in Sr$_2$IrO$_4$ and many others \cite{Crawford1994_214mag,Nakatsuji_CSRuO42000,Lee_OptRu2001,Mandrus_OsMIT2001,Cava_RuMIT1994,Shimura_Ir1195,cao2013frontiers}. These deviations were eventually put on a strong conceptual basis as the stabilization of a new ground state \cite{kim2009phase,kim2008novel}.

To understand this new phenomenology, one first must consider the orbital structure of the basic structural motif of the octahedral-coordinated iridates, Fig.~\ref{Jeff} -- an Ir $5d^5$ atom contained within an O octahedron. Assuming this octahedron has $O_h$ symmetry, the usual LS or Russell-Saunder's scheme relevant to $3d$ materials predicts a triply degenerate $t_{2g}$ orbital with a single hole in a low-spin configuration. Since the bandwidth of this $t_{2g}$ state is much larger than the small Hubbard $U$, one expects the material to be a simple band metal. However, this scheme has assumed negligible  \gls*{soc}. In iridates, both \gls*{soc} and crystal field splitting, Fig. \ref{Jeff}(a), have appreciable magnitude and their effects must be considered cooperatively. Introducing the \gls*{soc} effects to the $t_{2g}$ orbital scheme via jj-coupling, taken as $l_{\text{eff}} = 1$ states, yields a single \jeff{} orbital and a doubly degenerate $J_{\text{eff}} = 3/2$ orbital. Interestingly, the \jeff{} orbital is higher in energy, which can be understood as a consequence of the $L_{\text{eff}} = 1$ manifold being more than half full, or the $J=5/2$ shell being less then half-full, in accordance with Hund's third rule. Thus, we are left with a full $J_{\text{eff}} = 3/2$ quartet and half-filled \jeff{}  doublet.

The states derived from the \jeff{}  orbital have a strongly reduced bandwidth compared with the original $t_{2g}$ states, allowing the relatively weak coulomb repulsion, $\sim~1-2~\text{eV}$, to  generate upper and lower Hubbard bands with a finite gap [Fig.~\ref{Jeff} (b)]. This then stabilizes a novel local and isotropic valence configuration, Fig.~\ref{Jeff}(c), composed of an equal mixture of $xy$, $yz$, and $zx$ orbitals dubbed the \gls*{soc}-stabilized Mott state. The so-called isospin states are formed from a linear combination of the $t_{2g}$ orbitals with mixed up and down spins as $ | J_{\text{eff}} = 1/2,m_{J_{\text{eff}}} = \pm 1/2 \rangle= (|yz,m_s = \pm1/2\rangle\mp i|zx,m_s = \pm1/2\rangle \mp |xy,m_s = \mp1/2\rangle)/\sqrt{3}$ \cite{kim2008novel}. The equal mixing of the $t_{2g}$ orbitals requires perfect uniformity of the octahedra, which is, however, not exact in a real material; despite this, the \jeff{}  state often forms a good starting point for more in-depth consideration.

The total magnetic moment of this \gls*{soc}-stabilized Mott state is disparate from both the spin only $S = 1/2$ state seen with $3d$ orbitals and the atomic-like $J = 1/2$ state seen in rare-earths. In the former, the orbital moment is quenched and the total moment comes only from $\langle S_z\rangle$. In the latter, the total moment is $\langle L_z+2S_z\rangle = \pm1/3~(\mu_B)$, where the spin and orbital moments are anti-parallel with $|\langle S_z\rangle| = 1/6$ and $|\langle L_z\rangle| = 2/3$. In contrast, the \jeff{}  state is branched from the $J = 5/2$ manifold and therefore has $L_{\text{eff},~z} = -L_z$, giving  $\langle L_z+2S_z\rangle = \pm1$, while in practice the observed local moment is smaller owing to hybridization effects \cite{kim2008novel}.

\begin{figure}[t]
\includegraphics[width=1.0\columnwidth]{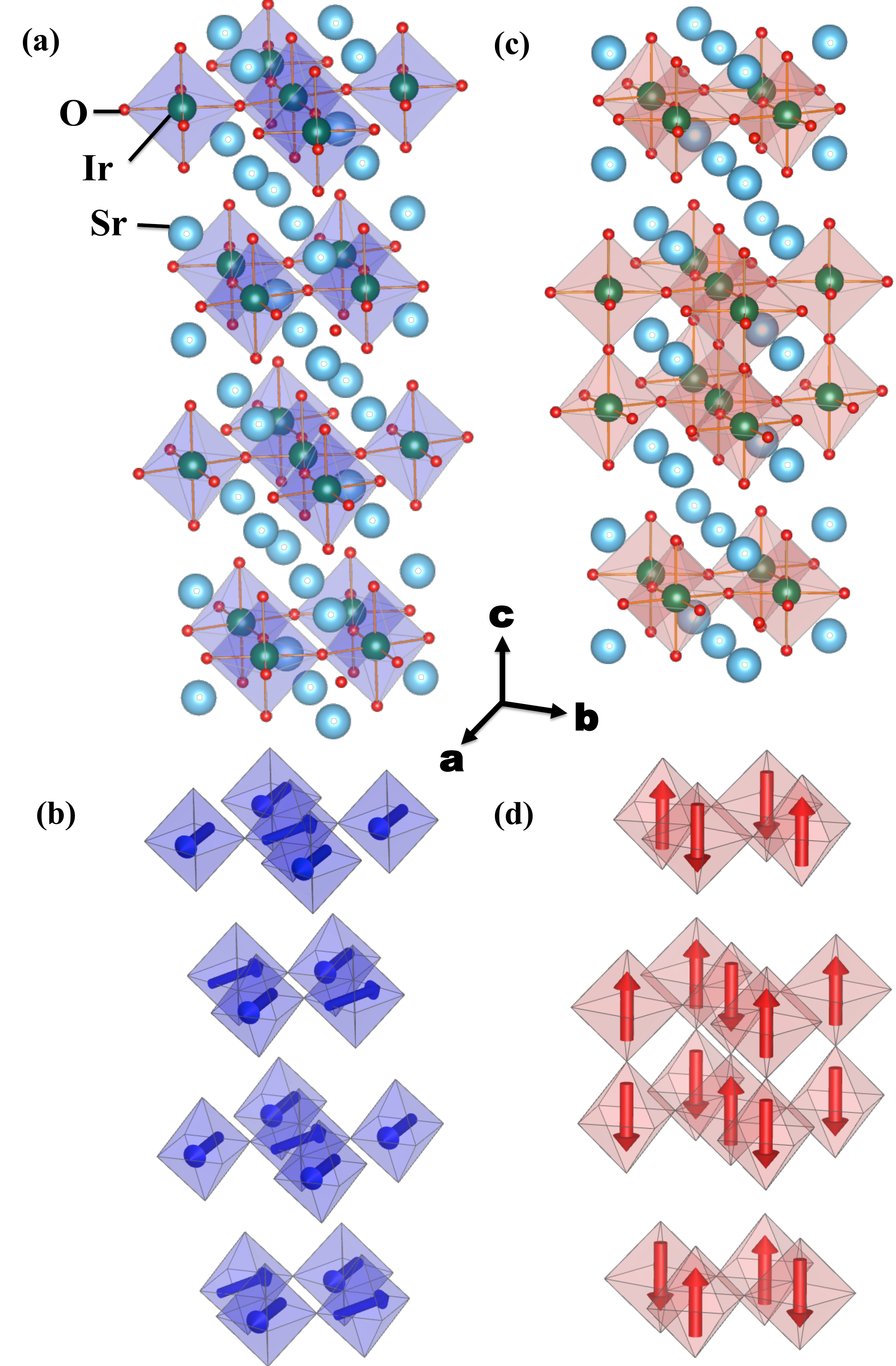}
\caption{\label{Mag_struc}  (a) Crystal and (b) $ab$-plane canted antiferromagnetic structure of \four{}, with ferromagnetic moment along $b$-axis \cite{boseggia2013_214}. (c) Crystal and (d) $c$-axis collinear antiferromagnetic structure found in \seven{} \cite{Boseggia2012_327REXS,Kim2012_327REXS}.
}
\end{figure}

These moments typically order antiferromagnetically, as superexchange dominates the magnetic interactions in corner-sharing octahedra iridate systems. Owing to the strong \gls*{soc}, two distinct insulating magnetic ground states have been observed in the Ruddlesden-Popper (RP) iridates. For isolated IrO$_2$ layers, as found in Sr$_2$IrO$_4$ and Ba$_2$IrO$_4$, the moment lies within the basal plane, Fig. \ref{Mag_struc}(a-b) \cite{boseggia2013_214,cao1998_214,Boseggia_B2IO4Mag2013}. While in Ba$_2$IrO$_4$ the in-plane Ir-O-Ir bond angle is $180^{\circ}$, in \four{} a large $c$-axis octahedral rotation is present changing the bond angle significantly. This distortion is especially important in $5d$ materials, due to the \gls*{dm} anisotropic magnetic interaction being of first-order in the \gls*{soc}. This coupling links the octahedral structural rotation with the \jeff{}  moment, leading to significant canting of the basal plane magnetic moments giving a small total ferromagnetic moment of $\sim0.1\mu_B$ observed in Sr$_2$IrO$_4$ but not Ba$_2$IrO$_4$ \cite{Crawford1994_214mag,kim2009phase,Moriya_Anis1960,dzyaloshinsky_thermodynamic1958}. In contrast to these two materials, for the bilayer iridate, \seven{}, interlayer Ir-O-Ir bonds are straight, forbidding the \gls*{dm} term by inversion symmetry,  and collinear moments along the $c$-axis are observed as shown in Fig. \ref{Mag_struc}(c-d) \cite{Cao2002_327structure,Boseggia2012_327REXS,Kim2012_327REXS}.
This dimensionality-driven change in the observed magnetic ground state is quite intriguing in light of the similar octahedral environments of the two systems. Employing \gls*{rixs}, it was shown that the $c$-axis N\'{e}el state is stabilized by the interlayer pseudodipolar anisotropic coupling that is of second order in the \gls*{soc} \cite{Kim2012_327RIXS}. As the dimensionality increases, a correlated paramagnetic metallic state is found in perovskite SrIrO$_3$ \cite{Longo_structure1971,Moon_PRL_2008,Liu2016_SIO}. Originally, this dimensional crossover was attributed to the three-dimensional (3D) structure which often increases the bandwidth and overcomes the Hubbard $U$. Subsequent work, however, has found this conventional picture is insufficient due to the strong SOC, as discussed in Sec. \ref{113_subseciton} \cite{Nie_ARPES_PRL}.

For the \jeff{} state in the single-layer iridates, many phenomenological parallels to the \gls*{htsc} $S = 1/2$ state of cuprates can be drawn. For instance, both host a lone electron or hole in the valence orbital, \gls*{afm} ordering with moments oriented in the CuO$_2$ (IrO$_2$) plane with finite canting, and very similar crystal structures. However, the different $t_{2g}$ vs $e_g$ character of the valence orbitals is an important caveat to this comparison. More detailed experimental and theoretical works have drawn more comprehensive comparisons, showing, for instance, similarly strong magnetic exchange interactions \cite{Kim2012_327RIXS,Coldea2001}, persistent magnetic excitations in the doped state \cite{Cao2017,Gretarsson2016_dop214RIXS,dean2013persistence,Meyers_LSCOmagnons2017}, similar electronic structure \cite{wang2011twisted}, and etc. Alluringly, evidence for close proximity to a possible \gls*{htsc} state has also been found in several works, making this system an excellent candidate for applying perturbations towards realizing this state \cite{gao_214SC_2015,yan2015electron,kim2016observation,Mitchell2015_214}. Beyond the possible applicability to \gls*{htsc}, iridates hosting edge sharing octahedra, among other structural motiffs beyond the scope of this review, are expected to host many exotic quantum phenomena, including spin liquid, topological Mott, Weyl semi-metal, etc. \cite{Witczak_correlatedSOC2014}.

In light of the desire to tractably alter the properties of the iridates towards these phases, a great deal of recent effort has focused on stabilizing thin films of the Ruddlesden-Popper iridates \cite{Liu2016_SIO,Lu_14_APL_113_strain, Nichols_13_APL_214_substrate, Lupascu2014strain214, Rayan_Serrao_PRB_2013, Domingo_15_Nano}. Application of epitaxial strain has been shown to allow controlled modulations to the material's properties \cite{Liu2016_SIO,Rayan_Serrao_PRB_2013,Oswaldo_theoryStrain2005,Dawber_FerroEleStrain2005,Chakhalian_ReviewSL2014,Hwang_ReviewSL2012}. Furthermore, layering of SrIrO$_3$ with band-insulating SrTiO$_3$ allows the creation of artificial analogues of the RP series iridates, with the advantage of enhanced strain control and the ability to synthesize structures not obtainable in bulk form \cite{Matsuno2015_SIOSTO,Hao_PRL_2017}. Finally, replacing the band-insulating layer with active 3d magnetic oxides creates interfacial magnetic interactions which can capitalize on the strong SOC of the 5$d$ state and introduce novel magnetic behavior \cite{Hirai_APLM_2015,Yi_arxiv_2017,moon_IrMn2017,Nichols_emerging2016,okamoto_charge2017}. Exploring the results of progress along these routes for altering the physical properties of the iridates forms the basis of this work.

In this review, we begin by detailing the synthesis techniques that allow the stabilization of the \gls*{rp} series iridates in thin film form. The application of this method to \four{} is then described, taking results of representative studies on strain manipulation and ultrafast laser excitation experiments as examples. We then move on to archetypal studies on films containing SrIrO$_3$, beginning with the effects of strain on the topological semimetallic state and then detailing the effects of interspacing these layers with band insulating and magnetically active layers. We end with an outlook on future possible directions of experimental and theoretical work in this rapidly expanding field. Further, beyond the corner-sharing octahedral iridates discussed here, great work has also been undertaken for several systems towards the realization of other quantum phenomena, e.g., Kitaev model spin liquids, topological states, etc \cite{Witczak_correlatedSOC2014,jackeli2009mott,Zhang_PyroIr2017,Kimchi_IrKitaev2014,Zhou_QSLreview2017}.

\section{Synthesis techniques}

Layered iridates of the \gls*{rp} phases can be thought of as a simple alternating stacking sequence of one SrO rocksalt layer and $n$ perovskite SrIrO$_3$ layers along the $c$-axis. The early members ($n = 1$ and $n = 2$) and the infinite end member were reported to be synthesized using conventional routes \cite{Crawford1994_214mag,kim2009phase, cao1998_214,subraman_MRB_1994,Boseggia2012_327REXS}. For example, perovskite SrIrO$_3$ has been stabilized under high pressure (albeit only  in  polycrystalline  form \cite{Longo_structure1971,Zhao2008_SIOstructure}), though bulk SrIrO$_3$ forms  the  6H  hexagonal  structure  at ambient conditions instead  of the perovskite phase. This thermodynamic instability has hindered investigations on perovskite  iridates,  but  can  be  solved  by  epitaxial  stabilization \cite{Jang_JPCM_2010,Nishio_16_Ir_growthdiagram,Jang_JKPS_2010}. Meanwhile, synthesis of bulk samples with $n$ in between is extremely challenging since these materials will naturally separate into mixtures of lower $n$ members and the $n = \infty$ phase, a general trend in \gls*{rp} materials \cite{Ruddlesden_AC_1957,Elcombe_ACSB_1991}.

\begin{figure}\vspace{-0pt}
\includegraphics[width=8cm]{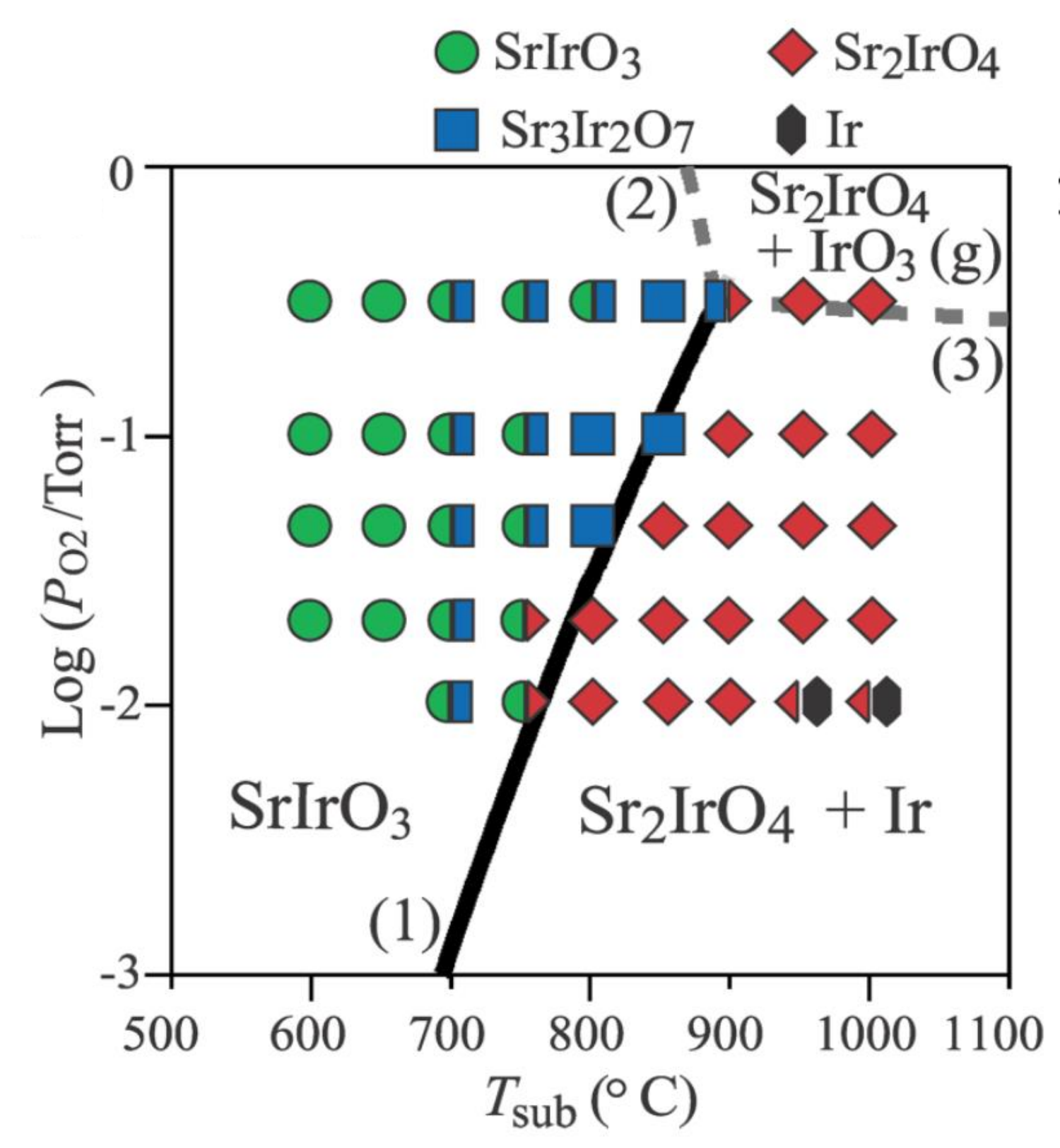}
\caption{\label{growth_phase} Growth phase diagram of partial oxygen pressure $P_{\text{O}_2}$ vs substrate temperature $T_{\text{sub}}$ for perovskite iridate films. Solid and dotted lines are calculated phase boundaries  from chemical equilibrium \cite{Nishio_16_Ir_growthdiagram}.
}
\end{figure}

To date, most of the known thin film deposition techniques have been applied to iridate \gls*{rp} phase growth, such as magnetron sputtering \cite{Fruchter_16_sputtering}, \gls*{pld} \cite{Kislin_113substrates, Horak_113, Zhang_15_PRB_LSAT, 113_liu_arxiv, Nishio_16_Ir_growthdiagram} and \gls*{mbe} \cite{Nie_ARPES_PRL,Liu_ARPES_SR}. In general, during a deposition process, the material of a target source first converts into vapor phase and then transports onto the surface of a substrate. The process of sputtering is realized through momentum transfer between accelerated positive ions and the target. Taking advantage of the plentiful amount of plasma, sputtering is well known for large-scale production. The deposition process is triggered through thermal evaporation in \gls*{mbe}. Featuring better control of vapor flow combined with in situ reflection high energy electron diffraction (RHEED), \gls*{mbe} enables thickness control at the atomic level. During a \gls*{pld} growth, a target is ablated by a pulsed laser with controllable fluence and frequency to reach a tunable deposition rate. This is crucial, especially for the growth of iridate films, due to the volatile nature of iridium at high temperature  \cite{Groenendijk_16_APL}. Additionally, when equipped with a RHEED system, \gls*{pld} can also realize thickness monitoring down to the atomic level. A detailed description of the other techniques can be found elsewhere \cite{Martin_review}.

Epitaxial SrIrO$_3$ films have been successfully realized on a series of substrates by means of \gls*{pld} and \gls*{mbe} growth. The growth has been recently reviewed in Ref. \cite{113_review_Biswas}. Contrary to SrIrO$_3$ films, growth of Sr$_2$IrO$_4$ films is usually more challenging due to its complex layer structure as mentioned before, which is common to many  \gls*{rp} oxides. Nevertheless, successful growth by \gls*{pld} and \gls*{mbe} has been reported in Refs. \cite{Liu2016_SIO,Lu_14_APL_113_strain, Nichols_13_APL_214_substrate,  Lupascu2014strain214, Rayan_Serrao_PRB_2013, Domingo_15_Nano,Lee_12_PRB_doping214}. Recently, the stabilization of Sr$_3$Ir$_2$O$_7$ film has also been achieved on SrTiO$_3$ substrates, by precise control of substrate temperature and oxygen partial pressure \cite{Nishio_16_Ir_growthdiagram}, as can be seen from the growth phase diagram of all three phases in Fig.~\ref{growth_phase}.

\section{Sr$_2$IrO$_4$ based systems}

The stabilization of the RP iridates through  thin-film growth techniques enables the study of numerous phenomena associated with the effect of the heterointerface and also facilitates experiments which are unfeasible in bulk single crystals. In this section, we discuss the findings for the influence of epitxial strain on \four{} and review one example of a ground-breaking experiment enabled by the availability of thin films.

\subsection{Strain-control over Ruddlesden-Popper iridate films}

As discussed in \ref{intro}, the \jeff{}  state is highly sensitive to deviations from isotropic octahedra, with some deviation nearly always being present in real materials. This sensitivity can be exploited towards either enhancing or reducing the \jeff{}  character, which is expected to have dramatic consequences for the physical properties. Specifically, the electronic gap, being set by the \gls*{soc}-stabilized Mott state, is expected to collapse upon straightening of the $Ir-O-Ir$ bond angle due to the increased bandwidth \cite{Moon_214Tdep2009}. Variation of transport behavior was indeed observed in \four{} films on (LaAlO$_3$)$_{0.3}$(SrAl$_{0.5}$Ta$_{0.5}$O$_3$)$_{0.7}$ (LSAT) substrates \cite{Lee_12_PRB_doping214} after chemical doping, which changes the bond angle and \gls*{soc} at the same time. Further, the magnetic ground state is also highly sensitive to the degree of tetragonal distortion, which is strongly dependent upon the local octahedral environment \cite{jackeli2009mott}.

\begin{figure}[h]\vspace{-0pt}
\includegraphics[width=1.0\columnwidth]{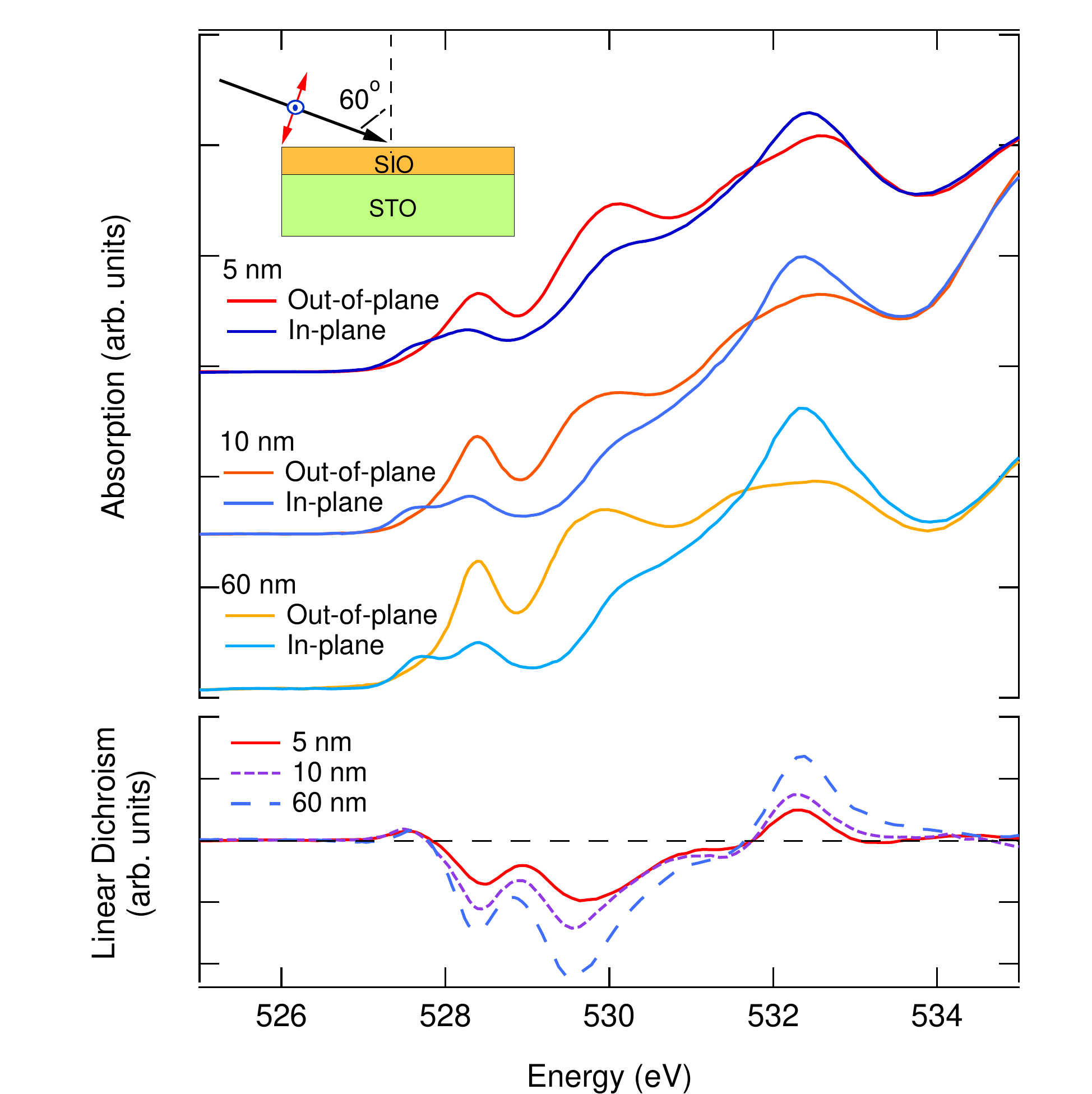}
\caption{\label{Serrao_XLD} Top panel: \gls*{xas} spectrum with horizontal and vertical polarization, as depicted in the inset, for each sample. Bottom panel: \gls*{xld} for each film showing recovery of \jeff{} leading to more uniform occupation of $t_{2g}$ orbitals. Applied in-plane strains are 0.17\%, 0.23\%, and 0.31\% for 60, 10, and 5 nm films respectively \cite{Rayan_Serrao_PRB_2013}.
}
\end{figure}

Along these lines, a well-established method for modulating the local octahedral environment in perovskites is the application of epitaxial strain, which was successfully utilized for films of iridates \cite{Liu2016_SIO,Lu_14_APL_113_strain,Nichols_13_APL_214_substrate,Lupascu2014strain214, Rayan_Serrao_PRB_2013, Domingo_15_Nano,Nichols_13_APL_214_orientation,miao_epitaxial2014,lu_214crossover2014}. In the work of Serrao \textit{et al.}, \four{} was deposited upon SrTiO$_3$ (001) substrates with varying thicknesses of 60, 10, and 5 nm \cite{Rayan_Serrao_PRB_2013}. By this methodology, various strains of 0.17\%, 0.23\%, and 0.31\% (-0.31\%, -0.59\%, and -1.40\%), with positive (negative) indicating tensile (compressive) strain, respectively, were surmised through diffraction to be applied in the $ab$-plane ($c$-axis).

\begin{figure}[h]\vspace{-0pt}
\includegraphics[width=1.0\columnwidth]{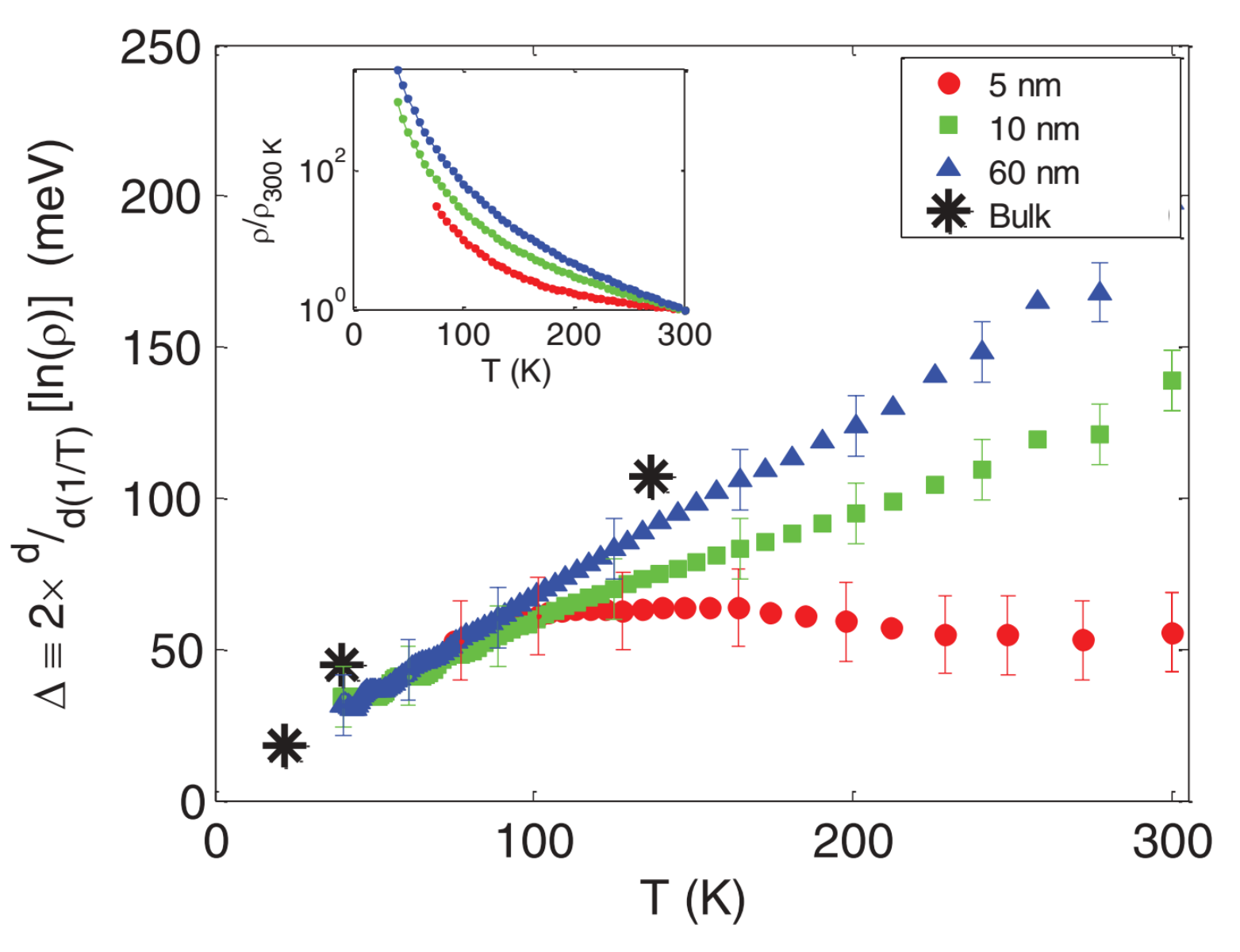}
\caption{\label{Serrao_trans} Thermal activation energy, $\delta$, as function of temperature, $T$, for each film and a bulk reference \cite{Ge_PRB_2011}. At room temperature the energy is strongly reduced as tensile strain is increased in the thinner films \cite{Rayan_Serrao_PRB_2013}. }
\end{figure}

Dramatic changes in the local octahedral environment were then evidenced by significant changes in \gls*{xas} measurements at the O $K$-edge with in- or out-of-plane x-ray polarization, Fig.~\ref{Serrao_XLD}. Here, the pre-peak feature from 527-529 eV comes from hybridization of the Ir $5d$ $t_{2g}$ and apical oxygen and planar oxygen $2p$ orbitals. With minimum applied strain, the local electronic anisotropy, evidenced by the spectra of the two polarizations and their difference spectrum shown in Fig.~\ref{Serrao_XLD}, closely resembles previous experiments on the bulk \cite{kim2008novel} where the internal \gls*{rp} structure imposes octahedral compression in the $ab$-plane. As the applied tensile strain increases and suppresses the internal compression, the observed electronic anisotropy in Fig.~\ref{Serrao_XLD}, bottom panel, significantly decreases, indicating enhanced strain-induced distortion of the octahedra, as expected, corroborating the modification of the \jeff{}  state. Modulation of the electronic properties was then deduced from changes in the electronic transport behavior.

Interestingly, room temperature resistivity was nearly identical for all applied strains, and transport measurements to low-$T$ showed all films displayed insulating behavior, inset of Fig. \ref{Serrao_trans}(a). However, despite these similarities there exists a clear difference in the slope of these curves, which evidences significant changes in the thermal excitation gap. Utilizing the thermally activated transport model, $\rho~\sim~e^{\delta/2T}$, a clear difference in the gap size at higher temperatures is observed, Fig. \ref{Serrao_trans}(a). The thick, nearly relaxed film has a 200~meV gap at room temperature, while the highest strain film has only 50~meV, a surprisingly large reduction. The gap size was previously reasoned to be mostly a function of the bonding angle, with a critical value of $\sim~170^{\circ}$, wherein the gap disappears, based upon calculations \cite{Moon_214Tdep2009}. However, for the strained films XRD refinement showed the upper bound on the bound angle of $\sim~160^{\circ}$, with theoretical results showing a bond angle change of $<~1^{\circ}$, hardly justifying the large change in thermal activation energy, Fig.~\ref{Serrao_trans}. Instead, the collapse of the $c$-axis (up to -1.4\%) leads to the favoring of the $xy$ orbital, which enhances the in-plane transport. Finally, the associated lessening of the Ir-O$_A$ bond also increases the hybridization further contributing to the lower gap, and corroborating the reduction of the local electronic anisotropy discussed previously. The strain effect on the electronic structure was also investigated in the work by Nichols \textit{et al.}\cite{Nichols_13_APL_214_substrate}, which grew Sr$_2$IrO$_4$ on four different substrates and found that the optical transition energy characteristic of the Hubbard $U$ increases from compressive to tensile strain. The width of the optical transition was, however, found to increase at the same time. As a result, the overall optical gap is rather unchanged down to the cutoff of the spectrum at $\sim$ 0.25 eV.

The rather large alteration of the thermal excitation gap with relatively small applied strain evidences the power of epitaxial strain to control the properties in \gls*{soc}-stabilized Mott systems, which are particularly sensitive to structural modulation \cite{Liu2016_SIO,Lu_14_APL_113_strain,Nichols_13_APL_214_substrate,Lupascu2014strain214, Rayan_Serrao_PRB_2013, Domingo_15_Nano,Nichols_13_APL_214_orientation,miao_epitaxial2014,lu_214crossover2014}. By way of comparison, nearly 30~GPa of external pressure was required to achieve similar changes within this system \cite{Haskel_214XMCD}.

Having established that the effects of applied strain on the electronic structure are quite dramatic, the strong \gls*{soc} implies strong perturbation of the magnetic behavior should also be expected. Indeed, theoretical calculations predict in the extreme case the magnetic ground state can be changed between in-plane canted and out-of-plane collinear orders with sufficient epitaxial strain \cite{jackeli2009mott,Kim2017_STSIOtheory,Kim2017_327theory}. While this phenomena is yet to be observed, the effect of strain on the magnetic behavior was still found to be significant \cite{Lupascu2014strain214}. Over a range of 1\% applied strain, spanning both compressive and tensile applied strain, a change in $T_{\text{N\'eel}}$ of 60K was observed. Further, utilizing \gls*{rixs}, a modulation of the magnetic exchange coupling was inferred from changes to the zone boundary magnetic excitation \cite{Lupascu2014strain214}. These findings display the sensitivity of the \jeff{} state to applied strain, with critical consequences for both electronic and magnetic behavior.

\subsection{Ultra-fast control of magnetic states with laser excitation}

Recent years have shown that ultra-fast laser excitation provides a compelling new tuning parameter to modify the behavior of correlated oxide based thin films and heterostructures  \cite{Zhang2014dynamics, Aoki2014,Dean2016, wall2016recent, Giannetti2016ultrafast, Gandolfi2017}. The need to match the pumped and probed sample volume, where the corresponding photons differ greatly in energy, is a particular challenge that films can help mitigate. Light pulses with different energy and polarization interact with materials in several different ways and can therefore be used to drive numerous different changes in materials. These include using optical energies to excite carriers from below to above the Fermi level \cite{Giannetti2016ultrafast} or using infra-red/terahetertz pulses to distort the lattice \cite{Zhang2014dynamics}. Within this field, iridate films occupy a special place, firstly because the Ir $L_3$ resonance opens new opportunities for probing magnetism in transient states and secondly due to the new channels of excitation afforded by \gls*{soc}.

\begin{figure*}\vspace{-0pt}
\includegraphics[width=1.3\columnwidth]{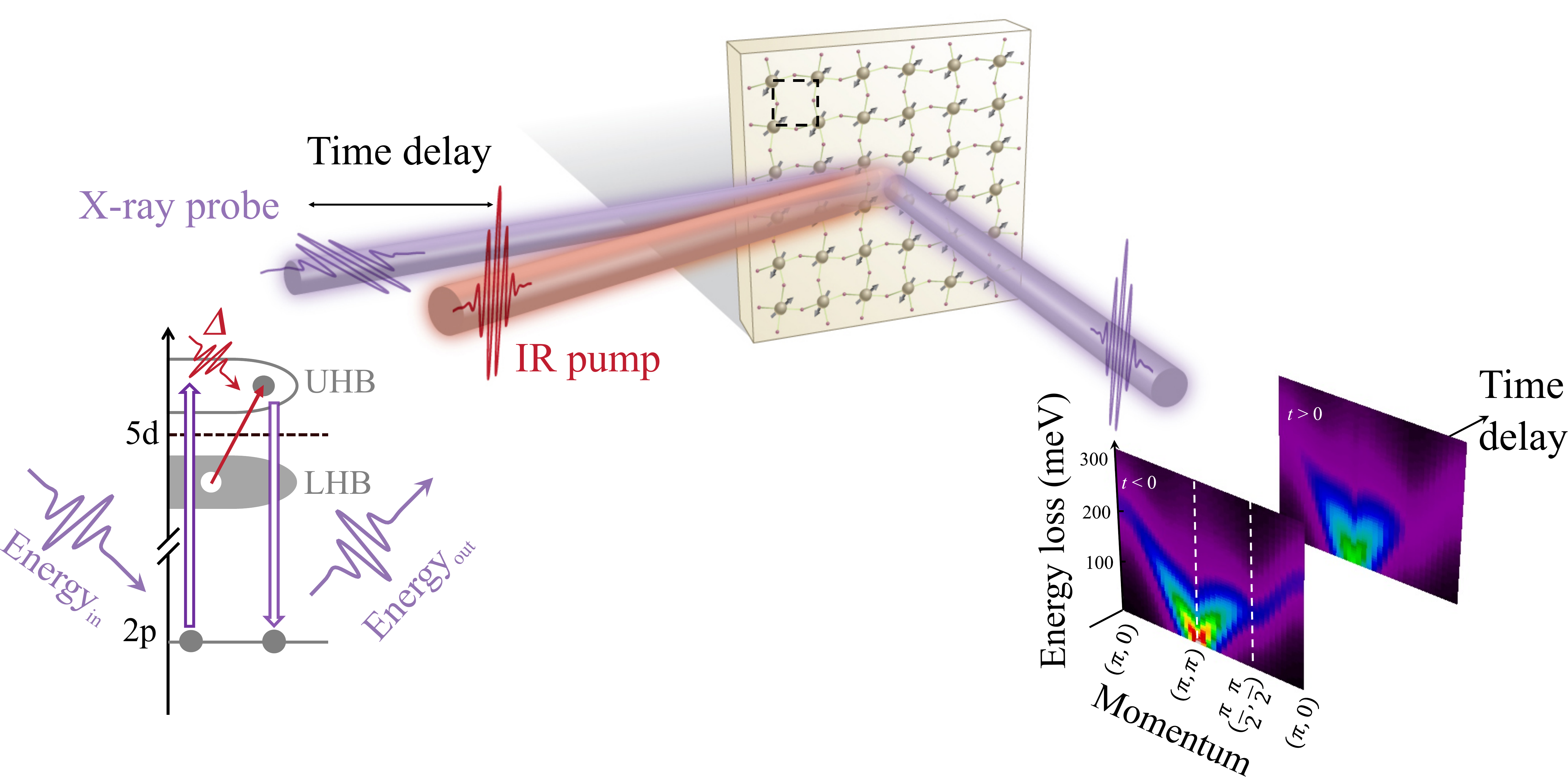}
\centering
\caption{\label{STO_TR_cartoon_Dean} An experimental setup for time resolved \gls*{rixs}. An infra-red beam shown in red excites Sr$_2$IrO$_4$ driving carriers across the Mott gap. An x-ray probe pulse, tuned to the Ir $L_3$-resonance, from a free electron laser then probes these states. The scattered x-rays are measured as a function of momentum, energy loss and time delay in order to probe the magnetic configuration in the transient state \cite{Dean2016}. }
\end{figure*}

Figure~\ref{STO_TR_cartoon_Dean} shows the setup used in a recent breakthrough experiment designed to ascertain the nature of magnetism in transient states. Sr$_2$IrO$_4$ was photo-doped and the resultant state was probed by performing \gls*{rixs} using a free electron laser \cite{Dean2016}. Laser-induced photo-doping has potential to create transient versions of the various exotic states that are accessible via chemical doping, with the advantage that the resulting states are tunable and reversible, while \gls*{rixs} provides a highly incisive probe of the magnetic quasiparticle spectrum \cite{dean2015insights}. This spectrum is a fundamental expression of the nature of the correlated electron state as it is the spatial and temporal Fourier transform of the spin-spin correlation function, and it encodes the interactions present in the magnetic Hamiltonian. Iridate films are very well suited to such experiments; as Ir sits in the 6th row of the periodic table its $L$-edge x-ray resonance occurs at a much higher energy than elements from the 4th row such as Cu   \cite{Lupascu2014strain214}. Ir x-ray resonant scattering can therefore easily access large momentum transfers in order to measure magnetic Bragg peaks. X-rays of this energy also propagate long distances in air, thus avoiding the requirements for ultra-high vacuum required for \gls*{rixs} studies of cuprates \cite{dean2015insights,sala2013high}.

\begin{figure}\vspace{-0pt}
\begin{center}
\includegraphics[width=0.8\columnwidth]{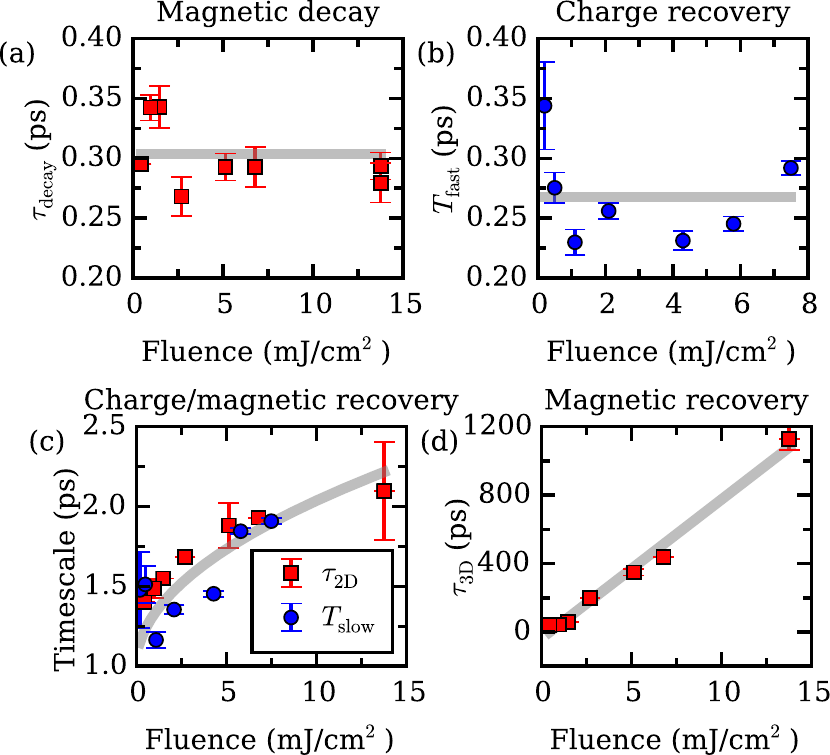}
\caption{\label{SIO_TR_timescales_Dean} A summary of the timescales governing magnetism (red squares) and charge (blue circles) in photo-doped Sr$_2$IrO$_4$. $\tau_\text{decay}$ is the timescale for the decay of magnetic correlations, $\tau_\text{2D}$ ($\tau_\text{3D}$) are the timescales for the recovery of two (three)-dimensional magnetism. $T_\text{fast}$/ $T_\text{slow}$ are the fast (slow) timescales for the recovery of the equilibrium charge distribution  \cite{Dean2016}.
}
\end{center}
\end{figure}

The magnetic dynamics of photo-doped Sr$_2$IrO$_4$ are summarized in Fig.~\ref{SIO_TR_timescales_Dean}. It was found that the out-of-equilibrium state, 2~ps after the photo-excitation, exhibits an almost complete suppression of long-range magnetic order provided the excitation fluence exceeds 6~mJ/cm$^2$ \cite{Dean2016,krupin2016ultrafast}. The recovery of these correlations was found to be highly anisotropic. Two-dimensional (2D) in-plane N\'{e}el correlations recover within a few ps; whereas the 3D long-range magnetic order restores on a fluence-dependent timescale of a several hundred ps. This was linked to the large difference in the in and out-of-plane magnetic exchange constants. The in-plane exchange is about 60~meV and the out-of-plane has been estimated to be on the order of 1~$\mathrm{\mu}$eV  \cite{kim2012magnetic, kim2014excitonic, Vale_PRB_2015, Fujiyama_PRL_2014}. The marked difference in these two timescales implies that the dimensionality of magnetic correlations is vital for our understanding of ultrafast magnetic dynamics. Going forward, there are many exciting opportunities to realize new physics in photoexcited iridates \cite{nembrini2016tracking, hsieh2012observation}. Of particular note is the fact that strong \gls*{soc} breaks the usual dipole selection rules expected for atomic transitions in iridates. Such an effect has already been exploited in O $K$-edge \gls*{rixs} studies of Sr$_2$IrO$_4$ \cite{liu_214singlemagnon2015}. These experiments probe magnons, excitations that are forbidden at the $K$-edges of light elements, but become allowed due to the exchange of orbital and spin angular momentum facilitated by \gls*{soc} in iridates \cite{liu_214singlemagnon2015, kim2015resonant}.

\section{SrIrO$_3$ based systems}

As compared to Sr$_2$IrO$_4$ with an alluring 2D magnetic structure, the other end member of the \gls*{rp} phases, perovskite SrIrO$_3$ is paramagnetic and actually famous for its exotic electronic structure \cite{Nie_ARPES_PRL,Zeb_PRB_2012,Carter_PRB_2012}. Nevertheless, the intrinsic strong \gls*{soc} in SrIrO$_3$ may manifest in introducing a novel magnetism when combined with other materials. Moreover, SrIrO$_3$ adopts a perovskite lattice structure when grown epitaxially, which can match with most of the well-known transition metal oxides, such as manganites. In this section, we will first introduce the strain engineering on the electronic structures of SrIrO$_3$ epitaxial films, and then list recent exciting findings on heterostructures composed of SrIrO$_3$ and other important materials.

\subsection{Topological semimetallicity in iridate films} \label{113_subseciton}

Earlier studies on bulk SrIrO$_3$ suggested the typical transport behavior of a paramagnetic metal at high temperatures \cite{Longo_structure1971,Zhao2008_SIOstructure}, and a low-temperature upturn attributed to a metal-to-insulator transition \cite{Zhao2008_SIOstructure}. Transport measurements on epitaxial thin film samples, however, indicated a semimetallic nature of the electronic ground state due to the small carrier density \cite{Jang_JPCM_2010,113_liu_arxiv}. The semimetallicity complicates the temperature-dependence of the resistivity, which may increase or decrease with decreasing temperature without invoking a true metal-to-insulator transition because both carrier density and mobility may significantly change upon thermal fluctuations. Moreover, semimetals usually have both electron- and hole-like Fermi surfaces, imposing additional complications to transport analysis. Nevertheless, the terms of "metallic" and "insulating" are often used in literature to describe the resistivity phenomenology \cite{Jang_JPCM_2010,Kislin_113substrates,Zhang_15_PRB_LSAT,biswas_JAP_2014,gruenewald_JMR_2014,Hirai_APL_2015,Wu_JPCM_2013}, which was found to sensitively depend on epitaxial strain and thickneses in film samples. The detailed mechanism of this behavior transition is still unclear.

\begin{figure}[t]\vspace{-0pt}
\includegraphics[width=8cm]{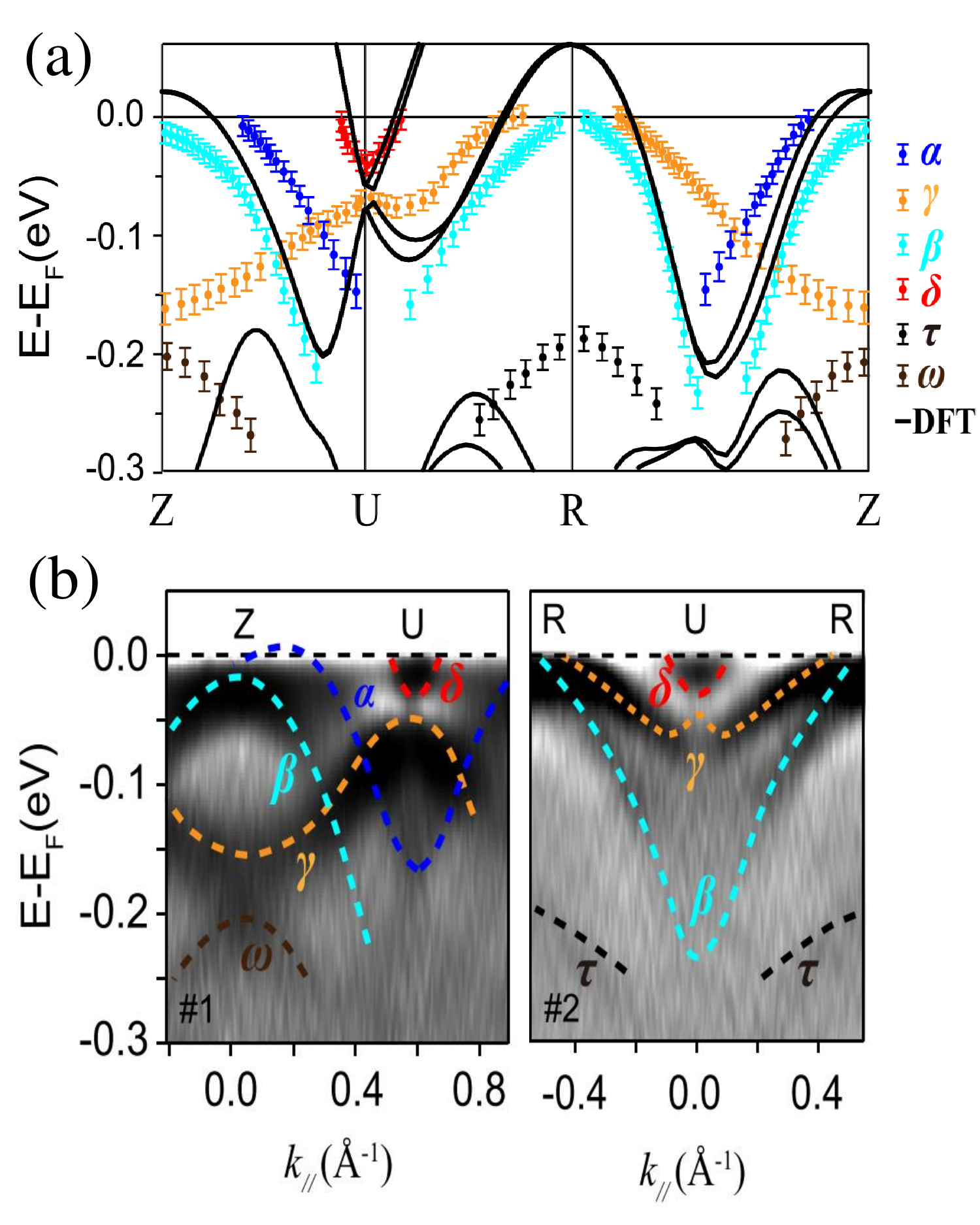}
\caption{\label{ARPES_113}(a) Experimental (symbols) and theoretical (line) band dispersions of SrIrO$_3$ film grown on SrTiO$_3$. (b) The second derivative images with respect to the energy along $Z-U$ (1) and $U-R$ (2) high-symmetry directions, confirmed a direct gap at $U$ points \cite{Liu_ARPES_SR}.}
\end{figure}

Theoretical study by Carter and Kee suggested not only a semimetallic state but also a nontrivial Dirac nodal ring around the $U$-point near the Fermi level \cite{Carter_PRB_2012}. Such a semimetallic band crossing prevents the Hubbard interaction from opening a charge gap, despite that most of the density of states are gaped away from the Fermi level \cite{Zeb_PRB_2012}. The semimetallic nature of the electronic ground state was demonstrated in two \gls*{arpes} studies. Namely, the bandwidth was found to be small and even narrower than Sr$_2$IrO$_4$, but without any charge gap \cite{Nie_ARPES_PRL}. Instead, a coexistence of hole-like and electron-like pockets were found  \cite{Nie_ARPES_PRL,Liu_ARPES_SR}. The former is caused by a hole-like flat band near the zone center, whereas the latter is associated with a linearly dispersing electron-band at the zone boundary. While the electron-pocket can be assigned to the upper Dirac cone, the crossing to the lower Dirac cone was not observed, i.e., a small gap is found between the $\delta$ and $\gamma$ bands shown in Fig.~\ref{ARPES_113} \cite{Liu_ARPES_SR}.

The absence of the band crossing was rather surprising because the Dirac nodal ring was originally thought to be associated with the mirror-symmetry of the $Pbnm$ space group \cite{Carter_PRB_2012}, which should be preserved even in a thin film structure. This result pointed to the need for resolving the exact crystal structure and space group in epitaxial films. In general, there are three main symmetry operations in addition to the inversion and time reversal symmetry in the $Pbnm$ space group, \textit{i.e.} $b$-glide, $n$-glide and mirror. Under a $b$-glide symmetry constraint, lattice structure is the same after reflecting in a plane perpendicular to the $a$-axis and translating along the $b$-vector. Similarly, $n$-glide symmetry allows crystal lattice to reflect in a plane perpendicular to the $b$-axis and translate along with a diagonal direction within the $ac$-plane. On the contrary, there is no translation operation in the mirror symmetry which is perpendicular to the $c$-axis. Therefore, without breaking translational symmetry or unit cell expansion, the mirror and glide symmetries can only be removed by breaking the reflection operation, \textit{i.e.} the angular lattice parameters.

To fully resolve the crystal structure in an epitaxial film, Liu \textit{et al.} have deposited SrIrO$_3$ films on orthorhombic (110)-oriented GdScO$_3$ substrates \cite{Liu2016_SIO}.
The refined lattice structure based on more than 70 structural Bragg peaks recorded using synchrotron x-ray diffraction found a deviation of $\gamma$ away from 90$^{\circ}$. Combining with dynamical diffraction theory calculations which fit the subtrate peaks, film peaks and their interference simultaneously, the new space group for the strained film was revealed as monoclinic $P112_{1}$/$m$, where the two glide symmetries are broken and the mirror is preserved.

\begin{figure}[h]\vspace{-0pt}
\includegraphics[width=8cm]{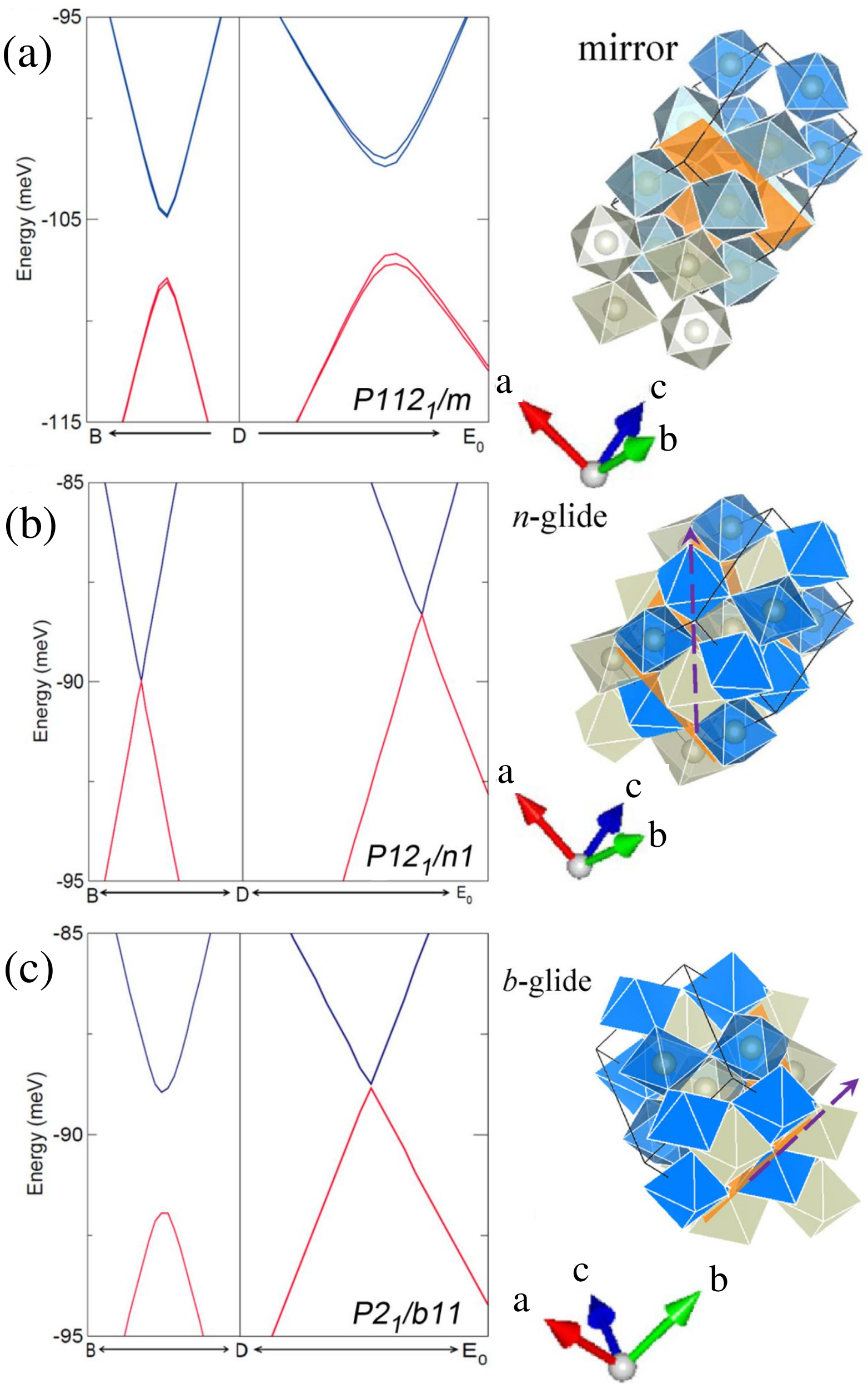}
\caption{\label{n-glide_liu_PRB} (a) The calculated electronic structure based on the experimental unit cell parameters and atomic positions. (b) and (c) show the electronic structures of a $P12$$_1$$/n1$ and $P2$$_1$$/b11$ structures, respectively. The right panels show the only preserved symmetries under the corresponding space group. The $D$, $B$, and $E_0$ points now equal to $U$, $X$, and $R$ points of the orthorhombic Brillouin zone \cite{Liu2016_SIO}.
}
\end{figure}

Resolving the lattice symmetry and atomic positions makes it possible to perform accurate band structure calculations using density functional theory. Interestingly, as shown in Fig.~\ref{n-glide_liu_PRB}(a), a clear gap opening around the $U$ point can be seen between the upper and lower Dirac cones, indicative of a lifted Dirac degeneracy. This was unexpected because the mirror symmetry is the only symmetry operation of the three that survived in $P112_{1}$/$m$, suggesting that the two glide operations might play important roles. Band structure calculations were further performed on another two  artificial structures with only the $n$-glide ($P12$$_1$$/n1$) and $b$-glide ($P2$$_1$$/b11$) symmetry preserved, respectively. In Fig.~\ref{n-glide_liu_PRB}(b), the Dirac nodal ring is fully persevered in the $P12$$_1$$/n1$ structure, revealing that the $n$-glide symmetry protects the Dirac degeneracy in addition to inversion and time-reversal symmetry. The Dirac nodal ring shrinks into a pair of Dirac points along the $D-E_0$ line ($U-R$ line in an orthorhombic zone) under $P2$$_1$$/b11$ [Fig.~\ref{n-glide_liu_PRB}(c)]. This observation suggests that, in the absence of the $n$-glide, the $b$-glide symmetry will take over to protect the Dirac degeneracy on the high-symmetry line of the $BZ$ boundary. Since both operations are glide planes, this symmetry protection renders SrIrO$_3$ a three-dimensional nonsymmorphic semimetal \cite{Fang_PRB_2015}.

This conclusion and the original study \cite{Carter_PRB_2012} by Carter and Kee can actually be well reconciled. It was proposed that the Dirac nodal ring will be turned into a pair of Dirac points in an artificial structure, Sr$_2$IrRhO$_6$, where the mirror symmetry is broken by alternating layers of Ir and Rh \cite{Carter_PRB_2012}. While the $b$-glide will be preserved, such a layer stacking would also remove the $n$-glide. The $n$-glide symmetry thus can be broken in multiple ways in epitaxial thin films and superlattices, highlighting the vital role of the symmetry identification in studies of SrIrO$_3$-based thin-film samples. This work paves another route to engineer the electronic structure for obtaining a novel topological phase \cite{Fang_NatureP_2016}, in addition to the geometric frustration as seen in iridates with edge-sharing IrO$_6$ octahedra \cite{Zhang_PyroIr2017,Kimchi_IrKitaev2014}.

\subsection{Toward 2D magnetism in artificial Ruddlesden-Popper iridates}

\begin{figure}\vspace{-0pt}
\includegraphics[width=8cm]{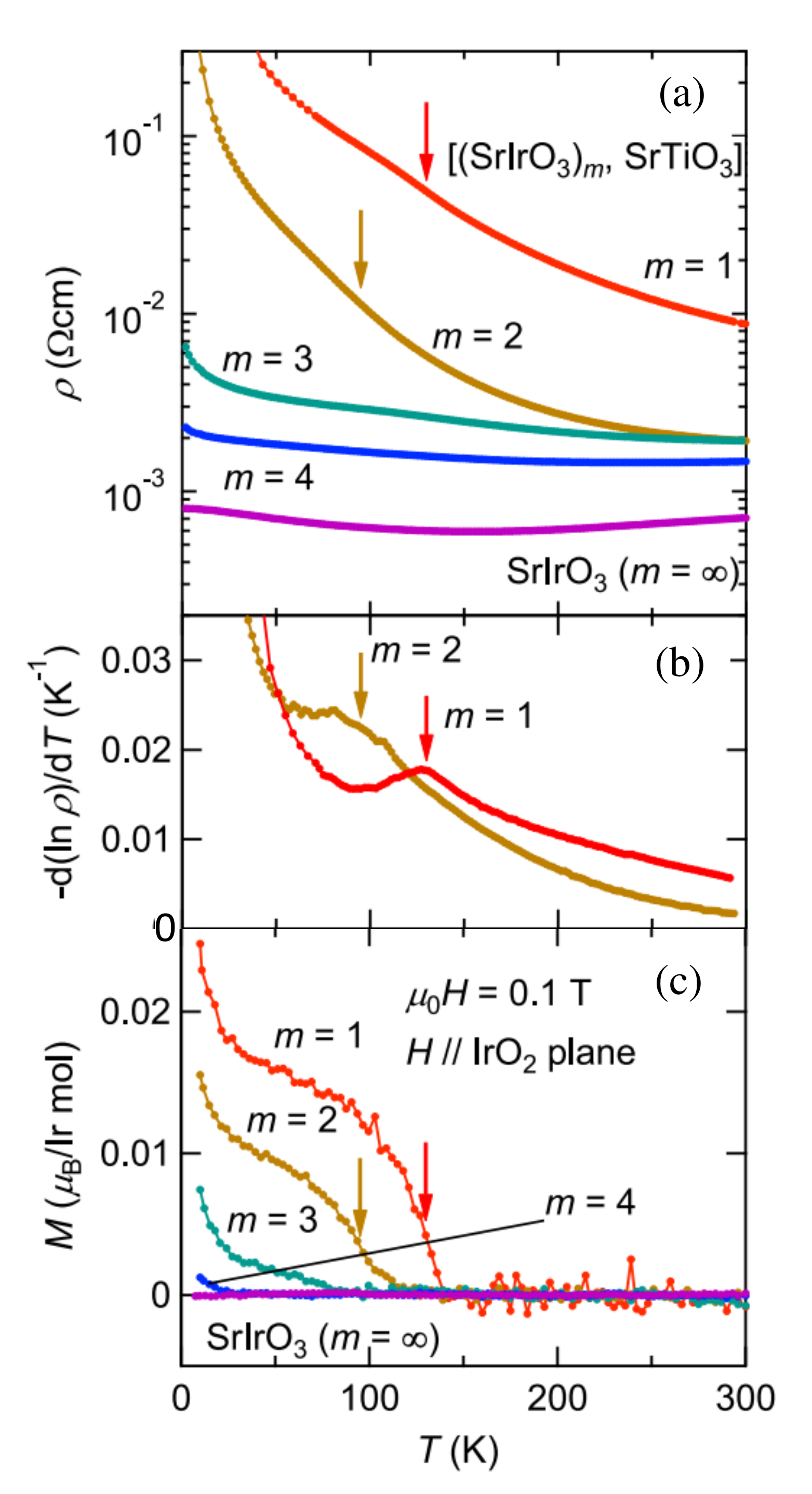}
\caption{\label{SISTO_Matsuno_PRL} Temperature dependence of resistivity (a) and  $-d(\ln\rho)/d(T)$ (b). The arrows indicate the temperatures of the observed anomalies. (c) In-plane magnetization of (SrIrO$_3$)$_m$/(SrTiO$_3$)$_1$ SLs as function of temperature. Arrows in panel (c) indicate the onset temperatures of magnetic ordering \cite{Matsuno2015_SIOSTO}.
}
\end{figure}

Dimensionality is another important effect that can significantly impact the electronic structure. An intuitive consideration is that dimensionality varies the number of the nearest-neighboring sites, which generally changes the bandwidth \cite{Moon_PRL_2008}. Experimentally, a dimensionality-controlled metal to insulator transition was indeed observed in the \gls*{rp} iridates. However, due to the nonsymmorphic semimetallic state in SrIrO$_3$, the dimensional crossover in layered iridates is, strictly speaking, a semimetal-to-insulator transition, where the dimensionality-driven confinement of the IrO$_2$ layers removes the semimetallicity and introduces a full gap. Additionally, considering the structural flexibility in an artificially designed lattice, epitaxial layering provides new avenues to further study and tailor such correlations \cite{kim2014electronic}.

For example, the layered \gls*{rp} structure was recently mimicked in an artificial superlattice (SL) \cite{Matsuno2015_SIOSTO}. In this pioneering work, the electronically and magnetically inert SrO monolayer in the unit cell of Sr$_{n+1}$Ir$_n$O$_{3n+1}$ was replaced with a monolayer of nonmagnetic dielectric SrTiO$_3$ in epitaxial superlattices of (SrIrO$_3$)$_m$/(SrTiO$_3$)$_1$. Using this strategy, the authors prepared not only the artificial counterparts of Sr$_2$IrO$_4$ and Sr$_3$Ir$_2$O$_7$, but also those of Sr$_4$Ir$_3$O$_{10}$ and Sr$_5$Ir$_4$O$_{13}$, which have never been reported in bulk synthesis. Transport measurements as shown in Fig.~\ref{SISTO_Matsuno_PRL}(a) revealed a semimetal to insulator transition upon decreasing the thickness of the SrIrO$_3$ layers, the same trend as that in bulk \gls*{rp} phases \cite{Moon_PRL_2008}. A monotonic increase of magnetization was also observed simultaneously in Fig.~\ref{SISTO_Matsuno_PRL}(c), indicating a strong coupling between electronic transport and magnetic moments. This coupling was further manifested in a resistivity anomaly originating from magnetic ordering in the $m = 1$ and $m = 2$ SLs [Fig.~\ref{SISTO_Matsuno_PRL}(b)].

Further evidence of the similarity between the superlattices and \gls*{rp} phases comes from their magnetic structures. Sr$_2$IrO$_4$ hosts an \gls*{afm} ground state and the directions of the magnetic moments are tied to the octahedral rotation due to the strong \gls*{soc} \cite{jackeli2009mott}. The octahedra rotation induces canting of the magnetic moments, which gives rise to a weak net moment in the $ab$-plane \cite{kim2008novel}. Similarly, a weak net magnetization was also observed in the $m = 1$ SL as seen from Fig.~\ref{SISTO_Matsuno_PRL}(c). Meanwhile, magnetic scattering investigation found magnetic Bragg peaks at (0.5 0.5 $L$) for the SL, demonstrated an \gls*{afm} configuration in the basal plane and \gls*{fm} coupling along the stacking direction. As compared to Sr$_2$IrO$_4$, the key difference is that, while the net moments of adjacent layers are spontaneously aligned with each other and lead to a macroscopic magnetization of the SL, they are anti-aligned in Sr$_2$IrO$_4$ and cancel each other \cite{kim2008novel}. This distinct interlayer coupling is likely due to the different stacking structure; while the adjacent layers have a (0.5, 0.5) lateral shift in Sr$_2$IrO$_4$, they are perfectly matched in the superlattices. Therefore, this work presents a new strategy to explore the physics of the dimensionality controllable metal-to-insulator transition.

\begin{figure}\vspace{-0pt}
\includegraphics[width=8cm]{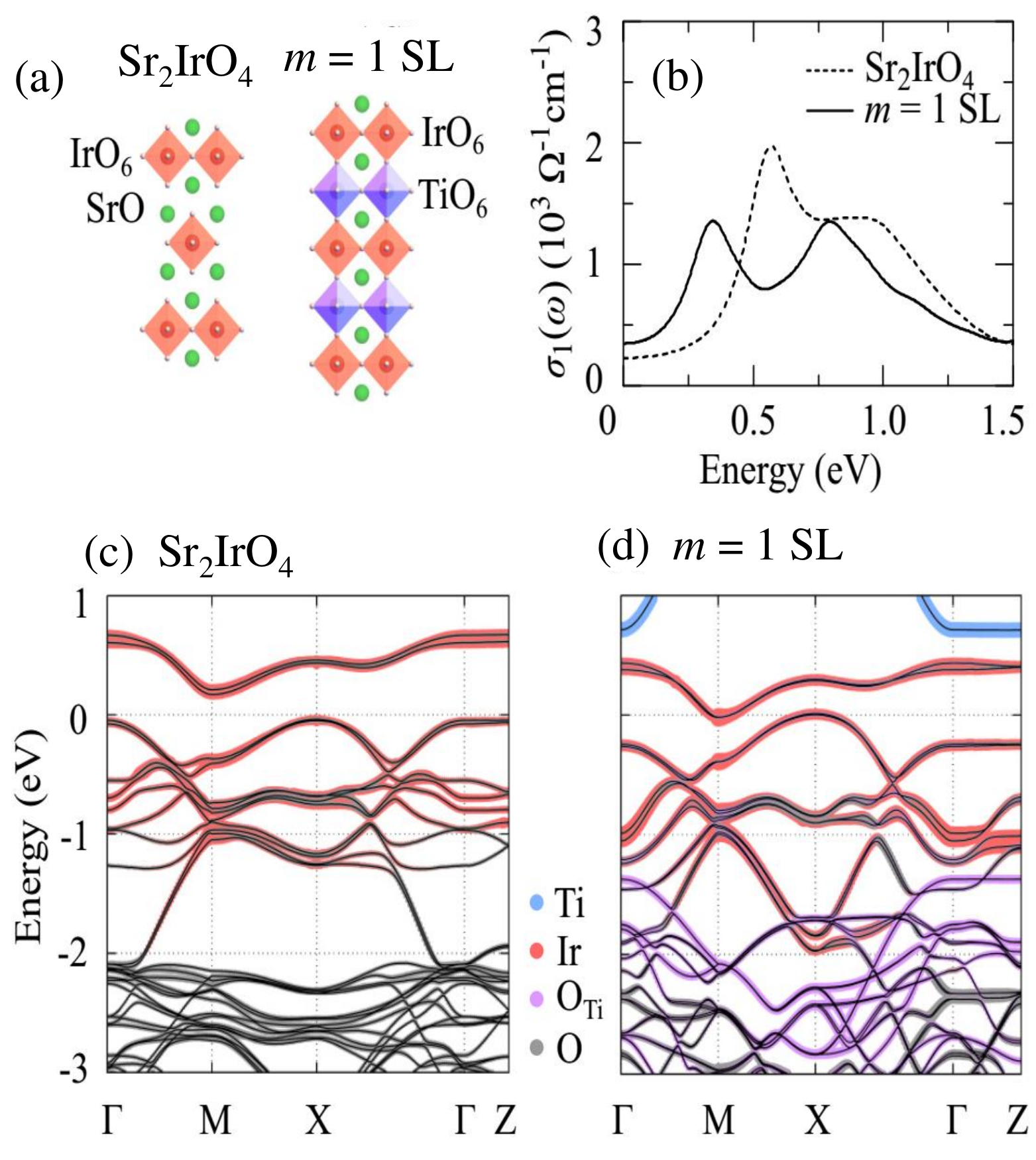}
\caption{\label{SISTOop_KIM_PRB} Schematic diagrams (a) and calculated optical conductivity, $\sigma_1(\omega)$, (b) of Sr$_2$IrO$_4$ and $m = 1$ SL. (c) and (d) displays the band structures of Sr$_2$IrO$_4$ and $m = 1$ SL, respectively \cite{Kim2016_OptDFT}.
}
\end{figure}

A subsequent optical study further confirms the similarity between the two series \cite{Kim2016_OptDFT}. However, red shifts were observed for the peaks characteristic of the transition from the lower to upper Hubbard bands in the SLs [Fig.~\ref{SISTOop_KIM_PRB}(b)]. Meanwhile, the low-energy spectral weight was also enhanced, indicating a reduction in the effective electron correlations in the SLs.  Density functional theory calculations verified the difference in the electronic structure. As shown in Fig.~\ref{SISTOop_KIM_PRB}(d), the $J_\text{eff} = 1/2$ bands of the $m = 1$ SL are more extended in comparison to that in Sr$_2$IrO$_4$ [Fig.~\ref{SISTOop_KIM_PRB}(c)], suggesting a larger bandwidth in the former. This result was ascribed to the additional Ir-Ir hopping channels through the SrTiO$_3$ spacer layers in SLs, as can be seen from the different atomic arrangements in the blocking layer between the quasi 2D layered structures, for example, Sr$_2$IrO$_4$ v.s. $m = 1$ SL [Fig.~\ref{SISTOop_KIM_PRB}(a)].

\begin{figure}[h]\vspace{-0pt}
\includegraphics[width=8cm]{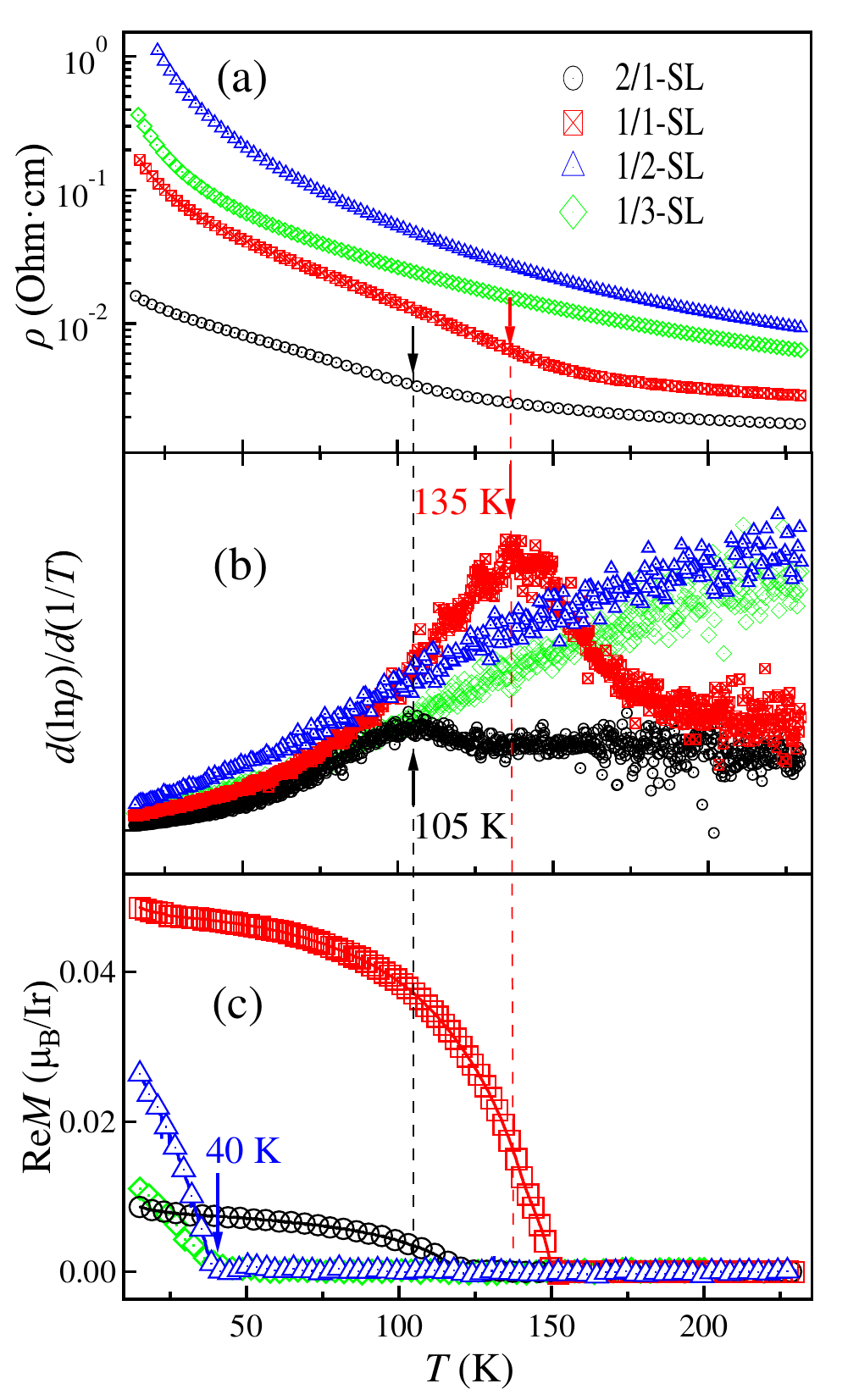}
\caption{\label{SISTO_R_Hao} Temperature dependence of in-plane resistivity (a) and $d(\ln\rho)/d(1/T)$ (b). The arrows denote the temperatures of the resistivity anomalies. (c) In-plane remnant magnetization as function of temperature. Before measurement in zero-field, samples were cooled down under 5 kOe in-plane magnetic field \cite{Hao_PRL_2017}.
}
\end{figure}

This result implies that details of the blocking layers provide extra flexibility in controlling physical properties of the confined 2D SrIrO$_3$ layers. To get a deeper insight, experimental realization of SLs with different SrTiO$_3$ blocking layer thicknesses, (SrIrO$_3$)$_m$/(SrTiO$_3$)$_n$ ((SrIrO$_3$)$_n$/(SrTiO$_3$)$_m$ was originally used), was recently achieved \cite{Hao_PRL_2017}. We use $m/n$-SL to denote these SLs. Inserting more SrTiO$_3$ blocking layer into the super unit cell is expected to separate neighboring SrIrO$_3$ layers and suppress the additional interlayer hopping channels in the superlattice structure, reinforcing the effective electron-electron correlations. Indeed, the insulating behavior manifested by the temperature-dependence of the $ab$-plane resistivity of 1/2- and 1/3-SLs is strengthened compared to 1/1-SL, as shown in Fig.~\ref{SISTO_R_Hao}(a).

More interestingly, the resistivity anomalies observed in 2/1- and 1/1-SL are invisible in 1/2- and 1/3-SLs with more than one SrTiO$_3$ layer in their super unit cell, as seen from the $\rho$-T plots [Fig.~\ref{SISTO_R_Hao}(a)] and also the $d(\ln\rho)/d(1/T)$ data [Fig.~\ref{SISTO_R_Hao}(b)]. Note that the resistivity anomaly coincides with the magnetic long-range order in $m$/1-SLs. Despite the absence of such an anomaly, the 1/2- and 1/3-SLs also show a clear magnetic phase transition at low temperatures [Fig.~\ref{SISTO_R_Hao}(c)], which indicates the coupling between transport and magnetism is weakened by the reduction of interlayer coupling.

The importance of interlayer coupling can also be seen from the much smaller onset temperature of the magnetic ordering in 1/2-SL (40~K) compared to that of 1/1-SL (150~K). The rapid decay of the ordering temperature when approaching the 2D limit indicates an exponential decrease of interlayer coupling with increasing blocking layer thickness. Further decrease of interlayer coupling does not significantly change the magnetic ordering stability, evidenced from the almost identical onset temperatures of 1/2- and 1/3-SLs, suggesting that the long-range magnetic order therein is maintained by spin anisotropy. Note that the magnetic interaction in the square lattice $J_\text{eff} = 1/2$ materials is dominated by the isotropic 2D Heisenberg term in the Hamiltonian \cite{jackeli2009mott,kim2012magnetic,Takayama_PRB_2016}, while long-range magnetic order in a 2D Heisenberg magnet is unstable at any non-zero temperatures according to the Mermin-Wagner theorem \cite{Mermin_PRL_1966} (La$_2$CuO$_4$ for example \cite{dean2012spin}). Therefore, the stabilization of the order in the 2D limit manifests the leading anisotropy term \cite{Cuccoli_PRB_2003,Cuccoli_PRL_2003}, which is led by the easy-plane anisotropy driven by the anisotropic exchange coupling \cite{jackeli2009mott}. In the absence of other anisotropy, the transition would be a Beresinskii-Kosterlitz-Thouless transition \cite{Berezi_JETP_1971,koster_JPCSSP_1973}; however, in this case, the small but finite corrections from higher-order terms, such as the compass-like term \cite{jackeli2009mott} and the residual interlayer coupling (see below) changes the transition into a second order phase transition of long-range order.

\begin{figure}[h]\vspace{-0pt}
\includegraphics[width=8cm]{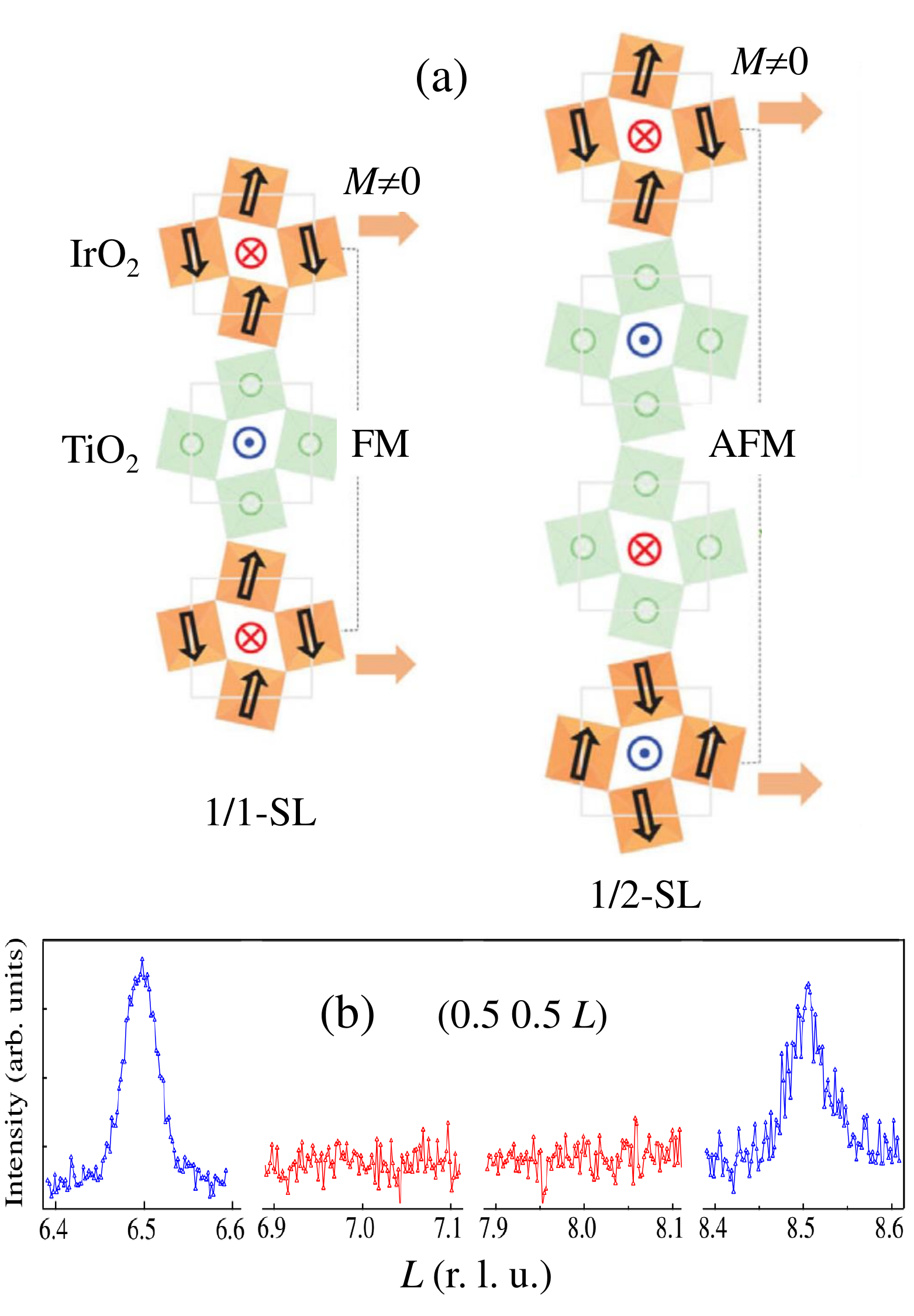}
\caption{\label{SISTO_M_Hao} (a) Magnetic structures and rotation patterns of 1/1- and 1/2-SLs. Black arrows indicate $J_\text{eff}$ moments, while orange arrows denote net moments in each SrIrO$_3$ layer. (b) $L$-dependence of peak intensity of (0.5 0.5 $L$) magnetic reflections. The measurement was performed at 10~K \cite{Hao_PRL_2017}.}
\end{figure}

Another intriguing observation is the macroscopic net magnetization in SLs with varied spacer thickness. For the 1/1-SL, a $c^-$ rotation occurs involving in-phase octahedral rotation of the adjacent SrIrO$_3$ layers around the $c$-axis \cite{Matsuno2015_SIOSTO}. Combined with a \gls*{fm} interlayer coupling, it aligns canted moments in each SrIrO$_3$ layer along the same direction [Fig.~\ref{SISTO_M_Hao}(a)]. However, under the same rotation pattern, the adjacent SrIrO$_3$ layers become out-of-phase in 1/2-SL. In this situation, a \gls*{fm} interlayer coupling would cancel out all the canted moments, contrary to the observed net magnetization. This can be reconciled if the interplayer coupling is \gls*{afm}. Such an anti-parallel alignment of the local moments between the adjacent layers, combined with the anti-phase rotation of the two layers, will lead to a parallel alignment of canted moment in each SrIrO$_3$ layer along the same direction, as can be seen from the Fig.~\ref{SISTO_M_Hao}(b). This hypothesis was confirmed through magnetic scattering. As shown in Fig.~\ref{SISTO_M_Hao}(c), magnetic peaks only appear at ${L + 1/2}$ in the 1/2-SL, where $L$ is an integer, rather than at ${L}$ as in the 1/1-SL. This result unambiguously shows that the interlayer coupling is \gls*{afm} in 1/2-SL opposite to the \gls*{fm} coupling in 1/1-SL. The comparison between their magnetic structures indicates that the interlayer coupling, although attenuated with thicker blocking layers, still plays a role in aligning the canted moments due to its variable sign commensurate with the switching phase relation of the octahedral rotation of the adjacent SrIrO$_3$ layers.

\subsection{Exploiting interfacial coupling between 5$d$ and 3$d$ states for novel magnetic states}

\begin{figure}[h!]\vspace{-0pt}
\includegraphics[width=7.5cm]{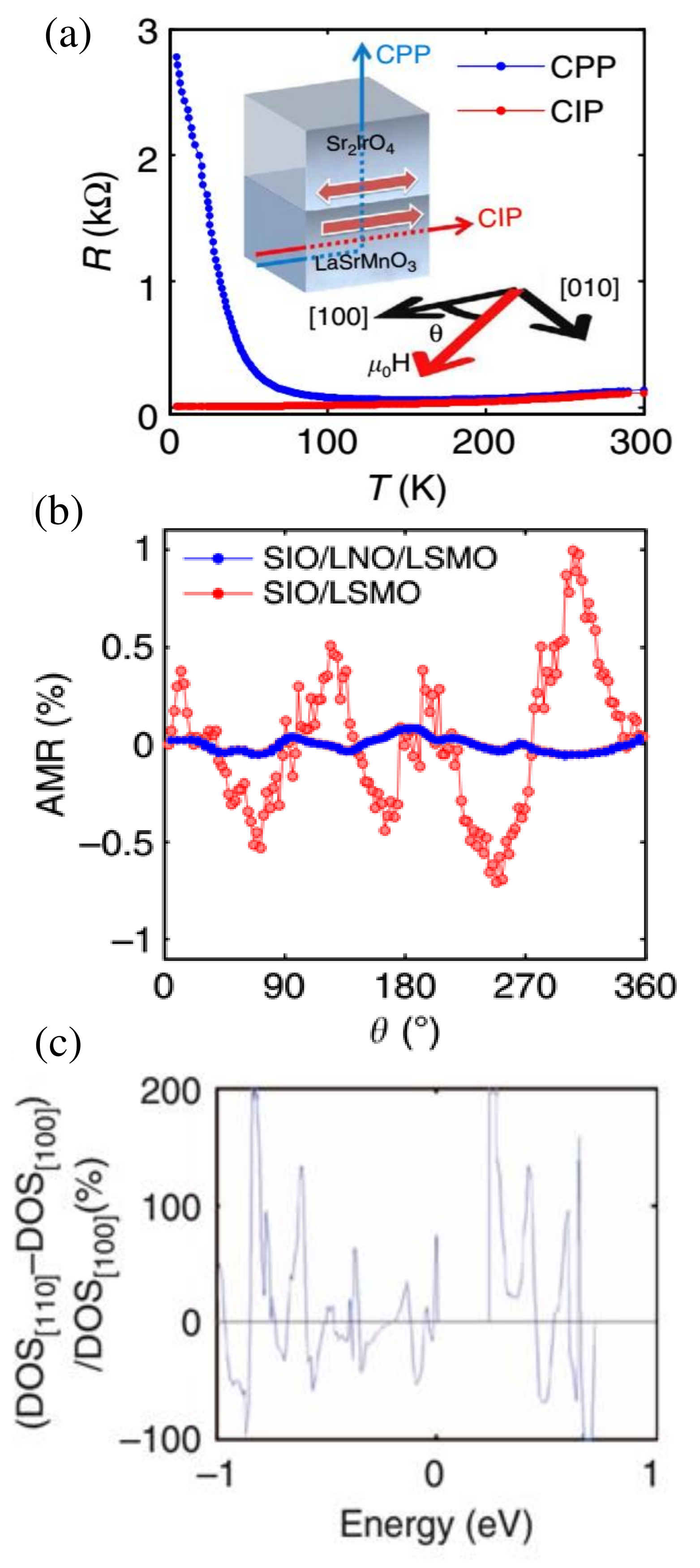}
\caption{\label{Natco_Fina_2014} Temperature dependent resistance of (a) La$_{2/3}$Sr$_{1/3}$MnO$_3$/Sr$_2$IrO$_4$ heterostructure. Inset shows the schematic diagram of current-perpendicular-to-plane and current-in-plane geometry. (b) Anisotropic magnetoresistance in \LSMO/Sr$_2$IrO$_4$ and \LSMO/LaNiO$_3$/Sr$_2$IrO$_4$ heterostructures measured at 4.2~K. (c) In-plane anisotropy of the density of states \cite{Fina_natcom_2014}.
}
\end{figure}

\begin{figure}\vspace{-0pt}
\includegraphics[width=7.5cm]{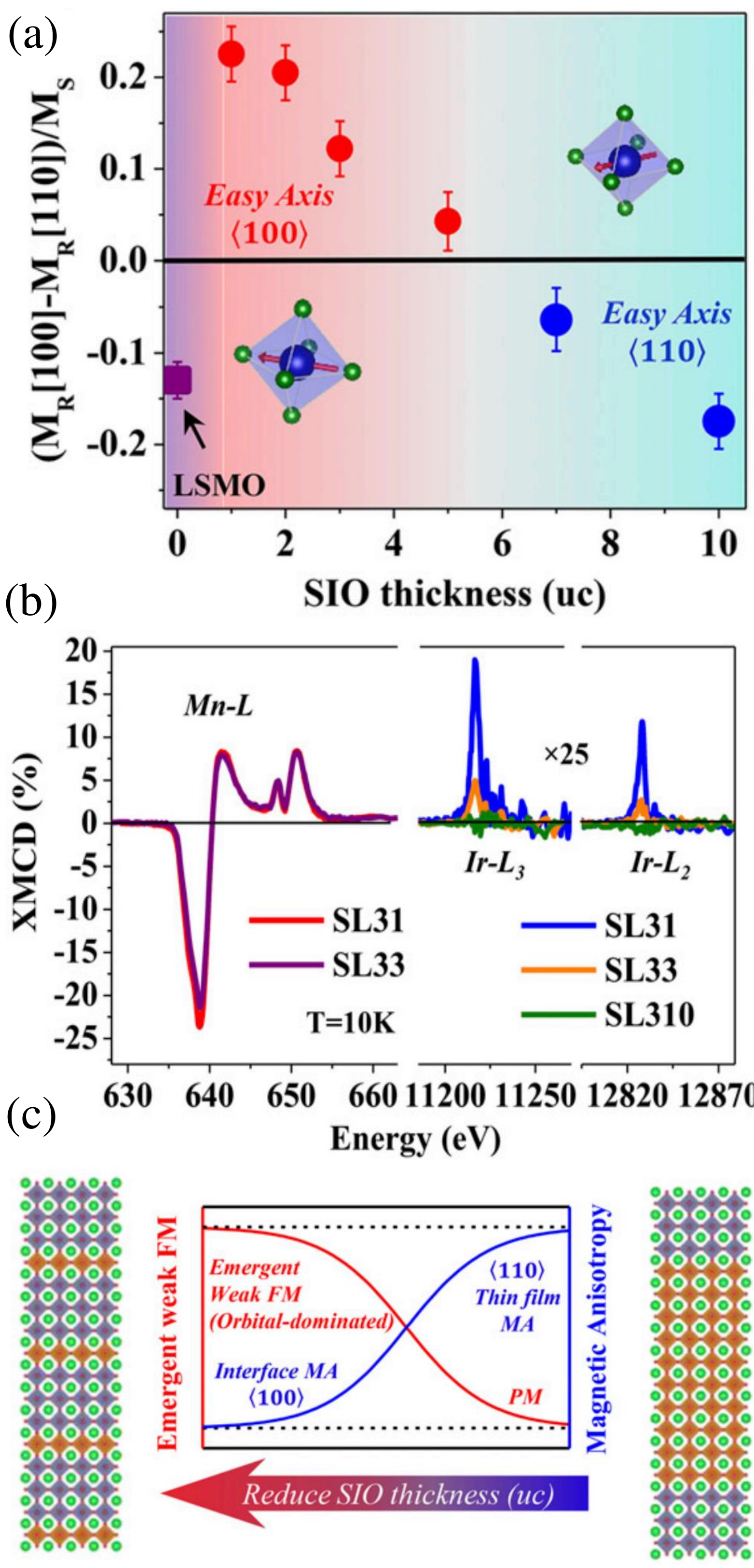}
\caption{\label{PNAS_Yi_2016} (a) Dependence of \gls*{ma} on SrIrO$_3$ layer number ($m$) in (\LSMO){$_3$}/(SrIrO$_3$)$_m$ (SL3$m$) supperlattices. Insets show the schematic diagrams of the two magnetic configurations. (b) Normalized \gls*{xmcd} spectra of a series of SLs. During measurement at 10~K, a 1~T magnetic field was applied along the $<100>$ direction. (c) Emergent ferromagnetism and anisotropy as functions of SrIrO$_3$ layer number based on theoretical calculations. \cite{Yi_PNAS_2016}.
}
\end{figure}

In $5d$ iridates, large \gls*{soc} entangles the charge and spin degrees of freedom, opening up the possibility of controlling the electronic properties through the magnetic state \cite{jackeli2009mott}. The $J_\text{eff}$ wavefunctions are encoded with spin components, crystal field orbital components, and the complex phases between different components. Change to any of these feature would affect the others and lead to different entangled effects. For instance, one may potentially rotate the spin-orbit moment relative to crystal axis and create significant changes to the density of states around the Fermi level. Such a rotation in an \gls*{afm} state requires rotating the \gls*{afm} axis, which is, however, difficult. One way to circumvent this issue is by dragging the moments through interfacial coupling, generally known as the exchange spring effect \cite{Morales_PRL_2015}. For example, Fina \textit{et al.} observed \gls*{amr} in a heterostructure comprising of Sr$_2$IrO$_4$ and FM La$_{2/3}$Sr$_{1/3}$MnO$_3$ \cite{Fina_natcom_2014}. The \gls*{amr} of Sr$_2$IrO$_4$ was found to have a four-fold symmetry as the magnetic field rotates the magnetization of the \gls*{fm} \LSMO, while there is no such \gls*{amr} by separating Sr$_2$IrO$_4$ through inserting a paramagnetic LaNiO$_3$ slab [Fig.~\ref{Natco_Fina_2014}(b)]. The origin of the observed \gls*{amr} was attributed to changes in the density of states near the band edges, above and below the charge gap, as the \gls*{afm} quantization axis of the spin-orbit wavefunction rotates within the 2D plane and senses the symmetry of the local environment. This effect can be measured by magnetoresistance because of the small band gap of Sr$_2$IrO$_4$, i.e., a narrow gap semiconductor. This study demonstrated the use of Sr$_2$IrO$_4$ as an \gls*{afm} semiconductor for novel spintronic applications. It is worthwhile to note that Fina \textit{et al.} used current perpendicular-to-plane geometry [Fig.~\ref{Natco_Fina_2014}(a)] in this study, taking advantage of the semiconducting nature of Sr$_2$IrO$_4$ \cite{cao1998_214}. Considering the in-plane \gls*{afm} configuration of semiconductor Sr$_2$IrO$_4$ \cite{kim2009phase,kim2008novel,Vale_PRB_2015,Takayama_PRB_2016,Fujiyama_PRL_2014}, there is no \gls*{amr} contribution due to varying the angle between the spin-axis and current. The observed \gls*{amr} is solely due to spin-axis rotation relative to the crystal axis \cite{Rushforth_PRL_2007}, originating from changes of the electronic structure induced by rotating the $J_\text{eff}$ moment. Similar behavior was also observed in a bulk crystal \cite{Wang_PRX_2014}. Further theoretical study verified this argument, \textit{i.e.} the density of states displays a strong sensitivity to the direction of the effective ${J}_{\rm eff}$ moment, as shown in Fig.~\ref{Natco_Fina_2014}(c). Note that though \gls*{amr} effect has been reported in devices composed of an \gls*{afm} metal electrode or a \gls*{fm} semiconductor \cite{Marti_NM_2014,Shick_PRB_2010}, from the application of view, the former is unsuitable for most of data processing electronics while synthesis of the latter with a high operating temperature is still technically challenging  \cite{Dietl_book_2008}.  This proof-of-concept study, thus, opens the possibility to integrate semiconducting and spintronic functionalities by utilizing AFMs.

\begin{figure}[h]\vspace{-0pt}
\includegraphics[width=8cm]{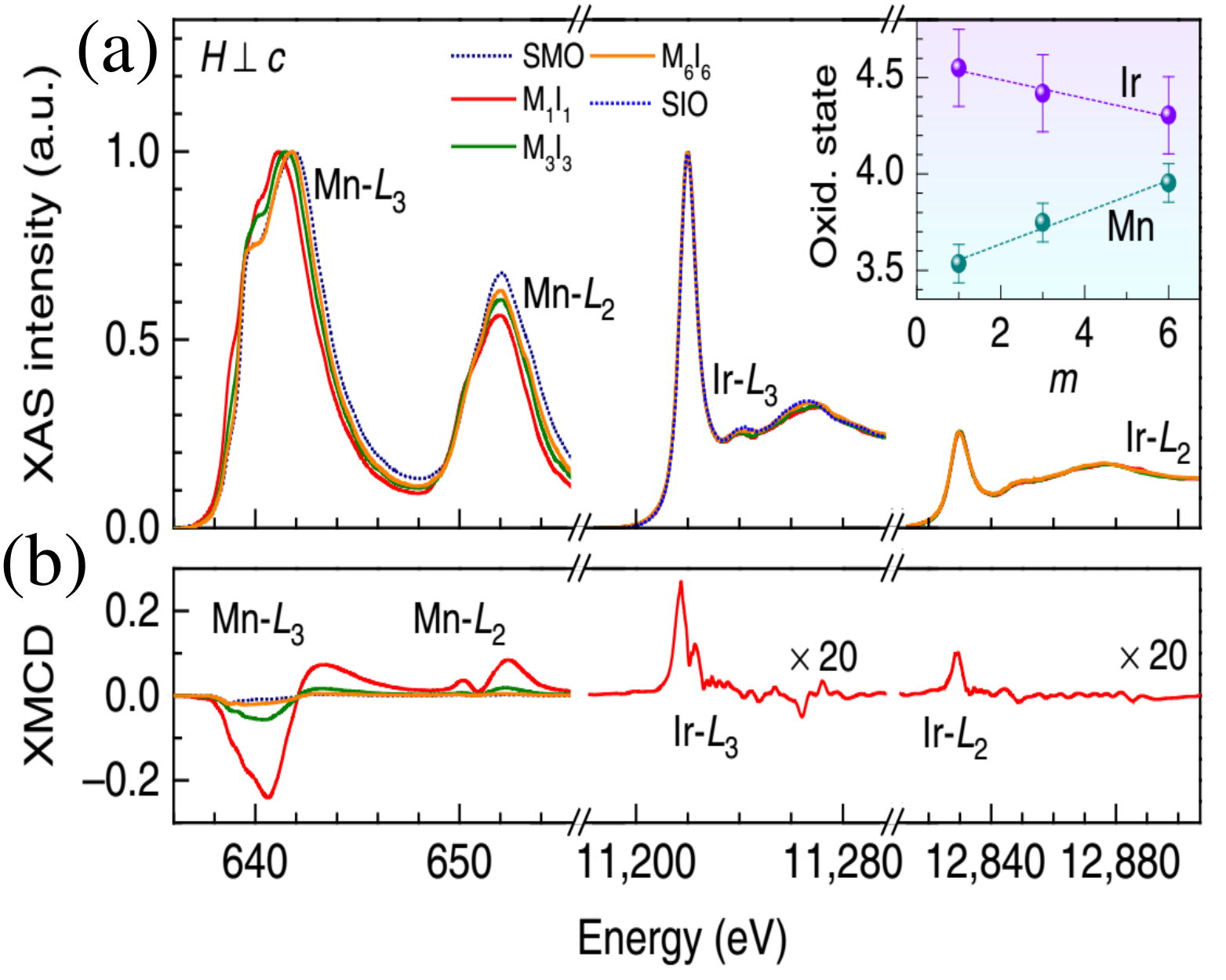}
\caption{\label{Natco_Nichols_2016} (a) \gls*{xas} spectra near the $L_3$ and $L_2$
edges of both Mn and Ir in superlattices composed of SrMnO$_3$ and SrIrO$_3$. Shifts of peaks can be seen in both elements indicating a charge transfer in between. Inset shows the extracted oxidation states of the two elements. (b) \gls*{xmcd} spectra of Mn and Ir $L$ edges. Finite intensity \gls*{xmcd} peaks revealed emergence of net magnetic moment in both elements. \cite{Nichols_natcom_2016}.
}
\end{figure}

While the \gls*{fm} La$_{2/3}$Sr$_{1/3}$MnO$_3$ layer was only used as a spin gate to drag the Ir spin-orbit moment through the interface, the interfacial coupling between 3$d$ and 5$d$ states is not a one-way street. The strongly spin-orbit-entangled state of 5$d$ electrons, while being highly susceptible to the coupling to the 3$d$ states across the interface, may also bring significant impact to the properties of the 3$d$ magnetic oxides \cite{Yin2013}. The first attempt toward this end was achieved in heterostructures comprising of \gls*{fm} \LSMO{} and paramagnetic SrIrO$_3$, taking the merits of the large spin moments in the former and the pronounced spin-orbit coupling in the latter \cite{Yi_PNAS_2016}. Though the overall magnetic behaviors of the heterostructures were governed by the \gls*{fm} \LSMO, its \gls*{ma} shows a systematic change with the number of SrIrO$_3$ monolayers. As shown in Fig.~\ref{PNAS_Yi_2016}(a), while the easy axis of \gls*{fm} \LSMO{} prefers to lie along the pseudocubic $<110>$ direction, inserting only one monolayer of SrIrO$_3$ per three monolayers of \LSMO{} switched the easy axis to $<100>$. As the SrIrO$_3$ layer thickness increases, the magnetic easy axis of \LSMO{} rotates systematically back to $<110>$, reaching a remarkable tunability \cite{Yi_PNAS_2016}. To investigate the original of this phenomenon, structural analysis was carried out and demonstrated that such a change of the \gls*{ma} is not due to shape anisotropy or symmetry changes. The possibility of charge transfer at interfaces was also excluded through \gls*{xas} measurements. Instead, \gls*{xmcd} measurements revealed the emergence of a weak \gls*{fm} moment in the nominal paramagnetic SrIrO$_3$ layer upon decreasing its thickness. Furthermore, there is a strong \gls*{xmcd} at the Ir $L_2$ edge [Fig.~\ref{PNAS_Yi_2016}(b)], indicative of a breakdown of the usual \jeff{} picture in the superlattices \cite{kim2008novel,Haskel_214XMCD}
. Indeed, sum rules analysis (also theoretical calculation) revealed a dramatic increase of the ratio between orbital moment and spin moment in the heterostructures compared to an ideal ${J}_{\text{eff}} = 1/2$ quantum state \cite{kim2009phase}. The enhanced orbital component features a mixture of ${J}_{\text{eff}} = 1/2$ and ${J}_{\text{eff}} = 3/2$ states \cite{jackeli2009mott}. Unlike the ${J}_{\text{eff}}$ = 1/2 state which has an isotropic orbital character \cite{kim2008novel}, the ${J_{\text{eff}} = 3/2}$ states are rather anisotropic. Their shapes resemble $t_{2g}$ crystal field orbitals but with strong spin-entanglement. In other words, the spin quantization axis strongly prefers to be aligned with crystal field quantization axis, \textit{i.e.} the Ir-O bond directions. This character locks the ${J}_{\text{eff}} = 3/2$ component to $<100>$, which is transfered via interfacial coupling to the spin moment in the \LSMO{} layers \cite{Yi_PNAS_2016}.

\begin{figure}[h]\vspace{-0pt}
\includegraphics[width=8cm]{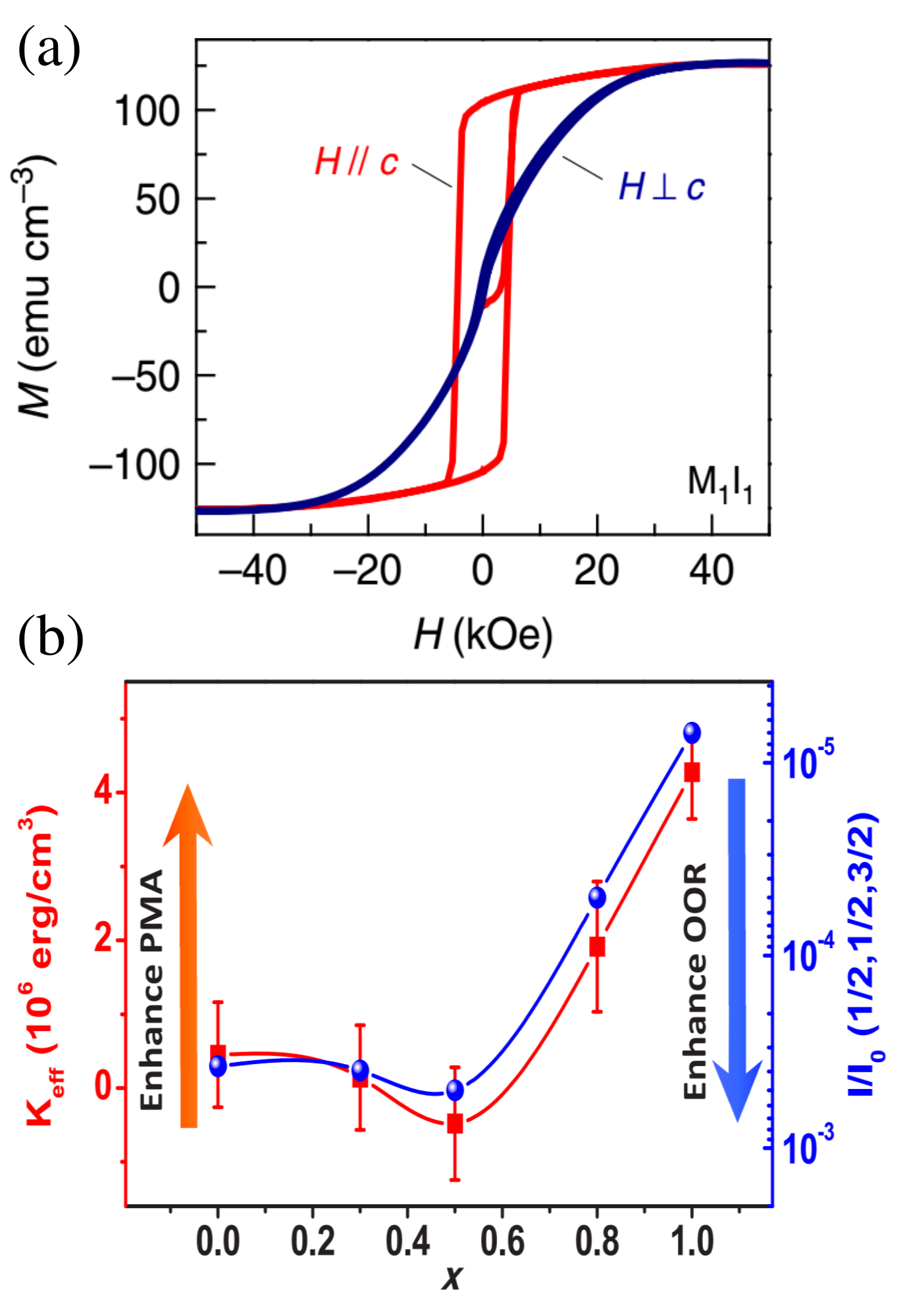}
\caption{\label{PMA_Yi_2017} (a) magnetic hysteresis loops of SrMnO$_3$/SrIrO$_3$ with magnetic field applied in-plane and out-of-plane \cite{Nichols_natcom_2016}. (b) Dependence of \gls*{pma} and \gls*{oor} on doping ratio $x$ of La$_{1-x}$Sr$_{x}$MnO$_{3}$/SrIrO$_3$ superlattices \cite{Yi_arxiv_2017}.
}
\end{figure}

While the easy-axis of the \gls*{fm} manganite remains in the plane of the film in the above study, rotation to the out-of-plane direction is possible when combined with charge transfer at interfaces \cite{Nichols_natcom_2016}. Nichols \textit{et al.} found various magnetic ground states in heterostructures composed of SrMnO$_3$, as well as SrIrO$_3$ slabs. Interestingly, although both Mn and Ir are in a 4+ valence state in each individual compound, electrons are transferred from Ir to Mn across interface when they are combined, turning on \gls*{fm} order in the nominally \gls*{afm} SrMnO$_3$. The interfacial charge transfer is revealed by \gls*{xas} as a blue shift of Mn $L_3$ edge and red-shift of Ir $L_3$ edge with increasing SrMnO$_3$, as well as SrIrO$_3$, slab thickness. These peak shifts indicate deviation of Mn and Ir valence states from their nominal values, as shown in the inset of Fig.~\ref{Natco_Nichols_2016}(a). Appreciable ferromagnetism was observed in heterostructures with only one SrMnO$_3$ and SrIrO$_3$ unit cells, where the effective charge modulation is strongest. This emerging ferromagnetism was shown to decay rapidly with increasing slab thickness.

More interestingly, a \gls*{pma} is realized in the \gls*{fm} SrMnO$_3$/SrIrO$_3$ heterostructure, deduced from the smaller coercive field in the out-of-plane magnetic hysteresis, as shown in Fig.~\ref{PMA_Yi_2017}(a) \cite{Nichols_natcom_2016}. Compared to the in-plane anisotropy in the \LSMO/SrIrO$_3$ heterostructure \cite{Yi_PNAS_2016}, this result suggests the ability to change the $A$-site cation in the manganite layer in order to rotate the magnetic easy axis. Controlling \gls*{pma} is an important step toward realizing magnetic storage devices. The \gls*{pma} also leads to an anomalous Hall effect \cite{Nichols_natcom_2016}. To shed light on the underlying mechanism, Yi \textit{et al.} recently performed a systematic study of the interfacial anisotropy by preparing a series of La$_{1-x}$Sr$_{x}$MnO$_{3}$/SrIrO$_3$ superlattices with doping ratio $x$ ranging from 0 to 1 \cite{Yi_arxiv_2017}. The \gls*{pma} was shown to emerge at $x = 0.5$ and then increases rapidly with $x$. The variation of \gls*{pma} was demonstrated can not be explained in modification of \gls*{ood}, which displays a linear dependence on doping ratio based on \gls*{xld} results. On the other hand, x-ray diffraction investigation revealed a similar dependence of \gls*{oor} on doping ratio [Fig.~\ref{PMA_Yi_2017}(b)], indicating a close correlation between the \gls*{oor} and \gls*{pma}. This result thus reveals the importance of the connectivity of oxygen octahedra in addition to the spin-orbit entanglement for tailoring interfacial functionalities.

\section{Outlook}

Recent research has uncovered a wealth of interesting new physics in the iridates derived from the interplay between electronic correlation and strong \gls*{soc}. The fact that even small structural modifications can impart large changes in electronic and magnetic behavior makes heterostructuring these materials an excellent route towards realizing new quantum phenomena. For example, tuning the tetragonal distortion and bond angle distortion under epitaxial strain has potential to realize quantum critical phenomena at the crossover between different antiferromagnetic states \cite{jackeli2009mott,Meyers_RIXS2017} as well as metal-to-insulator transition \cite{Kim2017_STSIOtheory}. The strong tie between the lattice degree of freedom and the electron and magnetic states highlights the importance of the thorough structural analysis in future investigations.

Another important area for future development is the realization of nontrivial topological states. Although the 3D topological semimetallic state protected by nonsymmorphic symmetries in ortho-perovskite iridates has been predicted in theoretical studies, direct experimental observation of the Dirac nodal ring has yet to be achieved. If it is experimentally verified, it would open the door to realizing a variety of topological electronic states that have been theoretically proposed \cite{Carter_PRB_2012,Fang_NatureP_2016, Chen_NatCom_2015,Chen_PRB_2014}. One may also realize 2D topological states in heterostructures by utilizing the strong \gls*{soc}. For instance, several theoretical proposals suggest that nontrivial band topology may emerge in bilayer SrIrO$_3$ grown along the cubic [111] direction \cite{Xiao_NatCom_2011,Wang_PRB_2011,Okamoto_PRL_2013,Rau_ARCMP_2016,Lado_PRB_2013}. The relatively strong electron-electron interaction in oxides also provide unique opportunities to investigate topological phases under electronic correlation \cite{Pesin_NatPhy_2010}. Towards this goal, the growth of perovskite iridates along the [111] direction was indeed recently realized \cite{Hirai_APLM_2015,Anderson_APL_2016}. Advances in combining atomic layering with photoemission spectroscopy and scanning tunnelling microscopy will boost the development in this area.

Finally, the exploration of the proximity effect and interfacial coupling of $5d$ states and $3d$ states has just begun. By engineering atomic stacking patterns, it should be possible to integrate the merits of these spin-orbit coupled oxides with other functional oxides. This includes realizing controllable \gls*{ma} in, for example, iridate-manganite heterostructures. An important future direction will be to utilize magnetic controls of this kind to achieve new switching mechanism in magnetic devices. Another fertile area involves studies of the topological Hall effect driven by artificially introducing \gls*{soc} at the interface of iridate and ruthenates \cite{Matsunoe_ScienceA_2016,Pang_ACSinterface_2017}. A major target of such work is the realization of an appreciably sized magnetic skyrmion. The wide variety of opportunities offered by interfacial charge transfer to modulate the electron-filling and obtain a desired quantum state without explicit dopants are only just being explored \cite{okamoto_charge2017, Nichols_natcom_2016}. Developments in theoretical calculations are also likely to play a key role in targeting the ideal heterostructures for the experimental realization of functional spin-orbit coupled interfaces.

\section*{Acknowledgements}
The authors acknowledge useful discussion with Neil J. Robinson and Yue Cao. J.L.\ acknowledges support from the Science Alliance Joint Directed Research \& Development Program and the Transdisciplinary Academy Program at the University of Tennessee. J.L.\ also acknowledges support by the DOD-DARPA under Grant No.\ HR0011-16-1-0005. M.P.M.D.\ and D. M.\ are supported by the U.S. Department of Energy, Office of Basic Energy Sciences, Early Career Award Program under Award No.\ 1047478. Work at Brookhaven National Laboratory was supported by the U.S.\ Department of Energy, Office of Science, Office of Basic Energy Sciences, under Contract No.\ DE-SC00112704.
\bibliography{refs}

\end{document}